\documentclass[aps,prd,superscriptaddress,preprintnumbers,floatfix,nofootinbib,notitlepage,showkeys,showpacs]{revtex4-1}
\pdfoutput=1
\usepackage[utf8x]{inputenc}
\usepackage{bbm}
\usepackage{cancel}
\usepackage{graphicx}
\usepackage{epsfig}
\usepackage{epstopdf} %This seems to be the package in cubbli
\usepackage{color}
\usepackage{slashed}
\usepackage{hyperref}
\usepackage{latexsym}
\usepackage{amsmath}
\usepackage{amsfonts}
\usepackage{amssymb}
\usepackage{color}
\usepackage{pdfsync}
\usepackage{epsfig}
\usepackage{epstopdf}
\usepackage{placeins}
\usepackage{subfigure}
\usepackage{comment}
\usepackage{slashed}

\usepackage[normalem]{ulem}

\newcommand{\Tr}{\textrm{Tr}\,}

\newcommand{\ev}[1]{\langle #1 \rangle}
\newcommand{\dd}{\textrm{d}}

\newcommand{\MSb}{\overline{\textrm{MS}}}

\newcommand{\csw}{c_{\rm{sw}}}

\newcommand{\be}{\begin{equation}}
\newcommand{\ee}{\end{equation}}
\newcommand{\bea}{\begin{eqnarray}} % only untightened
\newcommand{\eea}{\end{eqnarray}}

\newcommand{\bmp}{\noindent\begin{minipage}{16cm}}
\newcommand{\emp}{\end{minipage}\vskip 7mm} % 7mm untightened
\def\lsim{\mathrel{\raise.3ex\hbox{$<$\kern-.75em\lower1ex\hbox{$\sim$}}}}
\def\gsim{\mathrel{\raise.3ex\hbox{$>$\kern-.75em\lower1ex\hbox{$\sim$}}}}

\newcommand{\bs}[1]{{\bf #1}} %{\mbox{\boldmath{$#1$}}}

\newcommand{\gGF}{g_{\rm GF}}

\newcommand{\intron}[1]{}%{#1}
\begin{document}
\title{The gradient flow running coupling in SU(2) gauge theory with $N_f=8$ fundamental flavors}

%\author[a]{Viljami Leino}
%\author[b]{, Jarno Rantaharju}
%\author[a]{, Teemu Rantalaiho}
%\author[a]{, Kari Rummukainen}
%\author[a]{, Joni M. Suorsa}
%\author[a]{, Kimmo Tuominen}
%
%\affiliation[a]{Department of Physics and Helsinki Institute of Physics\\
% P.O.Box 64, FI-00014 University of Helsinki, Finland.}
% \affiliation[b]{Duke University\\
%Physics Bldg., Science Drive,  Box 90305, Durham, NC 27708, U.S.A}
%
%\emailAdd{viljami.leino@helsinki.fi}
%\emailAdd{jmr108@phy.duke.edu}
%\emailAdd{teemu.rantalaiho@helsinki.fi}
%\emailAdd{kari.rummukainen@helsinki.fi}
%\emailAdd{joni.suorsa@helsinki.fi}
%\emailAdd{kimmo.i.tuominen@helsinki.fi}

\author{Viljami Leino}
\email{viljami.leino@helsinki.fi}
\affiliation{Department of Physics, University of Helsinki \\
                      P.O.~Box 64, FI-00014, Helsinki, Finland}
\affiliation{Helsinki Institute of Physics, \\
                      P.O.~Box 64, FI-00014, Helsinki, Finland}

\author{Jarno Rantaharju}
\email{jmr108@phy.duke.edu}
\affiliation{Duke University \\
             Physics Bldg., Science Drive,  Box 90305, Durham, NC 27708, U.S.A}

\author{Teemu Rantalaiho}
\email{teemu.rantalaiho@helsinki.fi}
\affiliation{Department of Physics, University of Helsinki \\
                      P.O.~Box 64, FI-00014, Helsinki, Finland}
\affiliation{Helsinki Institute of Physics, \\
                      P.O.~Box 64, FI-00014, Helsinki, Finland}

\author{Kari Rummukainen}
\email{kari.rummukainen@helsinki.fi}
\affiliation{Department of Physics, University of Helsinki \\
                      P.O.~Box 64, FI-00014, Helsinki, Finland}
\affiliation{Helsinki Institute of Physics, \\
                      P.O.~Box 64, FI-00014, Helsinki, Finland}

\author{Joni Suorsa}
\email{joni.suorsa@helsinki.fi}
\affiliation{Department of Physics, University of Helsinki \\
                      P.O.~Box 64, FI-00014, Helsinki, Finland}
\affiliation{Helsinki Institute of Physics, \\
                      P.O.~Box 64, FI-00014, Helsinki, Finland}

\author{Kimmo Tuominen}
\email{kimmo.i.tuominen@helsinki.fi}
\affiliation{Department of Physics, University of Helsinki \\
            P.O.~Box 64, FI-00014, Helsinki, Finland}
\affiliation{Helsinki Institute of Physics, \\
            P.O.~Box 64, FI-00014, Helsinki, Finland}

%%%%%%%%%%%%%%%%%%%%%%%%%%%%%%%%%%%%%%%%%%%%%%%%%%%%%%%%%%%%%
\begin{abstract}% {%
We study the evolution of the coupling in SU(2) gauge field theory with
$N_f=8$ fundamental fermion flavors on the lattice.
This model is expected to have an infrared fixed point at high coupling.
We use HEX-smeared Wilson-clover action, and measure
the gradient flow running coupling with Dirichlet boundary conditions.
Extrapolating our results to continuum, we find
an infrared fixed point at $g_\ast^2 =8.24(59)_{-1.64}^{+0.97}$,
with statistical and systematic error estimates.
We also measure the
anomalous dimension of the quark mass operator,
and find its value at the fixed point $\gamma_\ast\simeq 0.15\pm 0.02$,
although for this quantity a reliable continuum limit is still lacking.
\end{abstract}

%\keywords{Lattice field theory, Conformal field theory}
\preprint{HIP-2016-33/TH}
%%%%%%%%%%%%%%%%%%%%%%%%%%%%%%%%%%%%%%%%%%%%%%%%%%%%%%%%%%%%%%%%%%%%%%%%%%%%%

\maketitle
%%%%%%%%%%%%%%%%%%%%%%%%%%%%%%%%%%%%%%%%%%%%%%%%%%%%%%%%%%%%%%%%%%%%%%%%%%%%%

\section{Introduction}

A SU($N$) gauge theory with $N_f$ massless flavors of Dirac fermions
transforming in the fundamental representation of the gauge group provides a simple probe of variety of gauge theory dynamics:
At small $N_f$ the theory breaks chiral symmetry of the vacuum similarly
as QCD, while above $N_f=11 N/2$ the theory is not asymptotically free.
Between $N_{f}^{(c)}\le N_f\le 11 N/2$, inside the so-called conformal window, 
the long distance behavior is expected to become governed by a nontrivial infrared stable fixed point (IRFP),
and the vacuum phase of the theory has infrared conformal behavior.

The determination of the location of the conformal window in a given gauge theory as a function of the numbers of colors,
flavors and fermion representations is interesting for our understanding of strong dynamics.
The theoretical value of $N_{f}^{(c)}$ can be estimated e.g. using the ladder approximation of
the Schwinger-Dyson equations for the fermion self-energy. These estimates suggest that the lower boundary is at
$N_f^{(c)}\simeq 4 N$~\cite{Appelquist:1986an, Sannino:2004qp}.
However, as the IRFP in typical cases is at strong coupling, nonperturbative methods are required,
and over recent years a lots of efforts in the field of lattice gauge theory has been devoted to address these questions;
see e.g. Ref.~\cite{Pica:2017gcb}.

In this paper we focus on SU(2) gauge theory.
In addition to gaining theoretical understanding on strong dynamics itself, this theory, at different values of $N_f$,
has applications for particle phenomenology beyond the Standard Model~\cite{Hill:2002ap,Sannino:2008ha,Sannino:2004qp}.
The effect of different numbers of flavors were systematically investigated in Ref.~\cite{Karavirta:2011zg}.
The $N_f=2$ case is a basic template for dynamical electroweak symmetry breaking~\cite{Hietanen:2014xca} and,
due to enhanced chiral symmetry,
this theory also has state which can act as novel dark matter candidate~\cite{Lewis:2011zb}.
The theories at larger $N_f$ are interesting since their renormalization flow can be very different from QCD-like theories
and they may serve as templates for walking technicolor.
At $N_f=10$ the existence of the fixed point has been demonstrated \cite{Karavirta:2011zg}, but at $N_f=8$ \cite{Ohki:2010sr} and
$N_f=6$ \cite{Karavirta:2011zg,Bursa:2010xn,Hayakawa:2013yfa,Appelquist:2013pqa}
the results are so far inconclusive.\footnote{%
However, preliminary results reported in Ref.~\cite{Leino:2016njf} indicate
the presence of the infrared fixed point at $N_f=6$.}

We complement the earlier results by providing analysis of SU(2) gauge theory with $N_f=8$ fermions in the fundamental representation.
We measure the running of the coupling constant using the gradient flow finite volume method~\cite{Fritzsch:2013je}
and establish the existence of an IRFP at $g_\ast^2 =8.24(59)_{-1.64}^{+0.97}$, with
statistical and systematical errors.
Preliminary results of these results have been previously reported
in Ref.~\cite{Rantaharju:2014ila,Suorsa:2015hoh,Leino:2015bfg,Suorsa:2016jsf}.

We also measure the anomalous dimension $\gamma$ of the fermion mass operator using two methods: the mass step scaling method and the Dirac operator spectral density.
The mass step scaling method works well at weak coupling, but near the infrared fixed point the lattice cutoff effects become uncontrollably large and a reliable result cannot be obtained.
On the other hand, the spectral density method remains stable at strong coupling,
and using our largest lattices we determine value of the anomalous dimension at the IRFP, $\gamma_\ast = 0.15\pm 0.02$.  However, for smaller lattice sizes the cutoff effects are too large, and thus a proper continuum limit of $\gamma_\ast$ is still lacking.

The paper is organized as follows:
In section~\ref{model} we define the model we study and outline the computational methods which we use.
The numerical results are presented in section~\ref{measurements}
and in section~\ref{conclusions} we present our conclusions and outlook for future work.

\section{The lattice implementation}
\label{model}
In this work we study the SU(2) gauge theory
with eight Dirac fermions in the fundamental representation of the gauge group.
In the lattice formulation we use the HEX-smeared~\cite{Capitani:2006ni} clover improved Wilson fermion action
with partially smeared plaquette gauge action;
\begin{equation}
  S=(1-c_g)S_G(U)+c_gS_G(V)+S_F(V),
  \label{eq:action}
\end{equation}
where $U$ and $V$ are, respectively, the unsmeared and smeared gauge link matrices.
$S_G$ is the standard single plaquette Wilson gauge action for the
SU(2) Yang Mills theory,
\begin{equation}
   S_G(U) = \beta_L \sum_{x;\mu<\nu}
   \left (1 - \frac12 \Tr [U_\mu(x) U_\nu(x+a\hat\mu)
  U^\dagger_\mu(x+a\hat\nu) U^\dagger_\nu(x) ] \right),
\end{equation}
where $\beta_L=4/g_0^2$, $c_g$ is the mixing parameter between the
smeared and unsmeared plaquettes,
and $a$ is the lattice spacing.
Using the partially smeared gauge action helps to avoid unphysical bulk
phase transitions at strong coupling~\cite{DeGrand:2011vp}.
We set the gauge action mixing parameter to the value $c_g=0.5$.
The detailed form of the smearing we use is described in Ref.~\cite{Rantaharju:2015yva}.

The clover improved Wilson fermion action is
\begin{equation}
  S_F
  = a^4\sum_{\alpha=1}^{N_f} \sum_x \left [
  \bar{\psi}_\alpha(x) ( i D + m_0 )
  \psi_\alpha(x)
   + a \csw \bar\psi_\alpha(x)\frac{i}{4}\sigma_{\mu\nu}
  F_{\mu\nu}(x)\psi_\alpha(x) \right ],
\end{equation}
where $D$ is the standard lattice Wilson-Dirac operator,
\begin{equation}
  D=\frac12 [\gamma_\mu(\nabla_\mu^\ast + \nabla_\mu ) -
  a\nabla_\mu^\ast \nabla_\mu],
\end{equation}
where $\nabla_\mu$ ($\nabla_\mu^\ast$) is the gauge covariant forward (backward)
lattice derivative using smeared link matrices:
\begin{equation}
  \nabla_\mu\psi(x) =\frac{1}{a}[V_\mu(x)\psi(x+a\hat\mu) - \psi(x)].
\end{equation}
The clover term contains the usual symmetrized field strength tensor,
and removes $O(a)$ errors from on-shell quantities with correctly
tuned Sheikholeslami-Wohlert coefficient $\csw$.
We use the tree-level value $\csw=1$,
which is a good approximation with smeared gauge links.

We use Dirichlet boundary conditions as in the Schrödinger functional method~\cite{Luscher:1992an,Luscher:1992ny,Luscher:1993gh,DellaMorte:2004bc},
with the gauge link matrices set to unity and fermion fields to zero at the temporal boundaries while the spatial boundaries are taken periodic:
\begin{align}
&U_k(0,{\bf{x}})=U_k(L,{\bf{x}})=V_k(0,{\bf{x}})=V_k(L,{\bf{x}})=1 \,,\nonumber\\%\;\text{when}\;x_0=0,L
&U_\mu(x_0,{\bf{x}}+L\hat{{\bf{k}}})=U_\mu(x_0,{\bf{x}})\,,\; V_\mu(x_0,{\bf{x}}+L\hat{{\bf{k}}})=V_\mu(x_0,{\bf{x}})\,,\nonumber\\
 &\psi(0,{\bf{x}})=\psi(L,{\bf{x}})=0\,,\; \psi(x_0,{\bf{x}}+L\hat{{\bf{k}}}) = \psi(x_0,{\bf{x}}) \,
\end{align}
where $k$ denotes coordinate in the spatial direction.
These boundary conditions facilitate the measurement of the mass anomalous dimension alongside the running coupling.
Furthermore, they remove the fermion zero modes and allow us to run simulations at vanishing physical quark masses.

The Wilson fermion action breaks the chiral symmetry explicitly and requires additive renormalization of the quark mass.
Thus we define $\kappa_c(\beta_L)$ as the value of the parameter $\kappa=1/(8+2 am_0)$
where the PCAC quark mass $aM(L/2)$, defined by the relation~\cite{Luscher:1996vw}, vanishes:
\begin{align}
  M(x_0) = \frac14 \frac{(\partial_0^\ast + \partial_0) f_A(x_0)}{f_P(x_0)}.
\label{eq:pcacm}
\end{align}
This relation receives an $O(a)$ improvement term~\cite{Luscher:1996vw},
but our use of the smeared gauge links renders its contribution very small and we
ignore it here.
The axial current and density correlation functions are:
\begin{align}
  f_A(x_0) &= %\frac{-a^6}{3L^6}
  -a^6 \sum_{\bs y,\bs z}
  \langle
    \bar\psi(x)\gamma_\mu\gamma_5 \lambda^a \psi(x)
    \,\bar\zeta(\bs y)\gamma_5\lambda^a\zeta(\bs z)
  \rangle
  \label{eq:fa}
  \\
  f_P(x_0) &= %\frac{-a^6}{3L^6}
  -a^6 \sum_{\bs y,\bs z}
   \langle
      \bar\psi(x)\gamma_5 \lambda^a \psi(x)
      \,\bar\zeta(\bs y)\gamma_5\lambda^a\zeta(\bs z)
   \rangle,
  \label{eq:fp}
\end{align}
where $\zeta$ and $\bar\zeta$ are boundary fermion sources at $x_0=0$, and
$\lambda^a$ is a fixed SU(8) generator acting on the flavor indices of the fermion fields.

To find the value of $\kappa_c$ we measure the mass at multiple values of $\kappa$ on lattices of size
$L/a=24$ and use interpolation to find the value where the PCAC mass vanishes.
We then use the same value of $\kappa_c$ on all lattice sizes.
On the largest volume $L/a=32$ this corresponds to a slightly negative mass of order $10^{-5}$.
The values of $\kappa_c$ used in the simulations are given in table~\ref{table:kappa}.
We see no indication of bulk phase transitions even at strongest couplings (smallest $\beta_L$) used.

\subsection{Measurement of the coupling}
\label{sec:couplingdefs}

We measure the running of the coupling using the Yang-Mills gradient flow~\cite{Narayanan:2006rf,Luscher:2009eq}
combined with the Schrödinger functional finite-volume scaling~\cite{Fritzsch:2013je}.
To set up this method, we introduce an extra coordinate, flow time $t$
and a flow gauge field $B_{\mu}(x;t)$.
The flow field $B_{\mu}$ evolves according to the flow equation
\begin{align}
  \partial_tB_{\mu}&=D_{\nu}G_{\nu\mu} \,,\;
\end{align}
where $G_{\mu\nu}(x;t)$ is the field strength of the flow field $B_{\mu}$ and $D_\mu=\partial_\mu+[B_\mu,\,\cdot\,]$.
The initial condition is defined in terms of the original continuum gauge field $A_\mu$ such that $B_{\mu}(x;t=0) =A_\mu(x)$.

To leading order in perturbation theory in SU($N$) gauge theory, the field strength evolves as~\cite{Luscher:2010iy}
\begin{align}
\langle E(t) \rangle &= -\frac{1}{4} \langle G_{\mu\nu} G_{\mu\nu} \rangle(t) = \frac{3(N^2-1)g_0^2}{128\pi^2 t^2}+\mathcal{O}(g_0^4).
\label{eq:edisc}
\end{align}
The flow smooths the gauge field over the radius $r\sim \sqrt{8t}$,
systematically removing the UV divergences
and automatically renormalizing gauge invariant observables~\cite{Luscher:2011bx}.
Thus, the flow can be used to define the coupling at scale
$\mu=1/\sqrt{8t}$ nonperturbatively as
\begin{equation}
\label{eq:g2gf}
\gGF^2(\mu)=\frac{128 \pi^2}{3(N^2-1)} t^2\langle E(t)\rangle\big\vert_{t=1/8\mu^2}\,,
\end{equation}
which agrees with perturbation theory to the leading order.

In the lattice formulation we consider the case $N=2$ and set up the flow equation on the lattice.
The continuum flow field is replaced by the lattice link variables $U_{\mu}(x;t)$ which are evolved according to
\begin{equation}
\label{eq:linkflow}
\frac{\partial}{\partial t} U_{\mu}(x;t) = -g_0^2
\left(\frac{\partial}{\partial U_\mu(x;t)} S_{\rm{GF}}[U]\right)U_{\mu}(x;t)
\end{equation}
with the initial condition $U_{\mu}(x;t=0) = U_\mu (x)$.
For the flow evolution action $S_{\rm{GF}}$ we use the tree-level improved L\"uscher-Weisz pure gauge theory action~\cite{Luscher:1984xn}.
We measure both symmetric clover and simple plaquette discretized observables for $\langle E(t)\rangle$.
Unless otherwise indicated we use the clover discretization in our analysis.

In order to limit the scale into a regime $1/L \ll \mu \ll 1/a$,
where Eq.~\eqref{eq:g2gf} is free of both lattice artifacts and finite volume effects,
we relate the lattice scale to the renormalization scale by defining a dimensionless parameter $c_t$
as described in Refs.~\cite{Fodor:2012td,Fodor:2012qh,Fritzsch:2013je}:
\begin{equation}
\label{eq:flowscale}
\mu^{-1} = c_tL = \sqrt{8t}.
\end{equation}
A range of $c_t=0.3-0.5$ is suggested in Ref.~\cite{Fritzsch:2013je} for the SF scheme.
Within this range the cutoff effects, which are minimized at $c_t=0.5$,
statistical variance, and boundary effects~\cite{Luscher:2014kea},
both of which grow with the $c_t$, are reasonably small.

Unless otherwise specified, we use $c_t=0.4$ in our analysis,
but we also compare with other values of $c_t$.
In order to minimize the effects of the fixed SF boundaries at $x_0 = 0$ and $L$,
we measure the expectation value of the gauge field energy~\eqref{eq:edisc} only on the central time slice $x_0 = L/2$.

Since we are not using perfectly improved observables and actions in our flow~\cite{Ramos:2015baa},
the gradient flow coupling $\gGF^2$ will have cutoff effects.
In order to minimize these cutoff effects at the continuum limit, we add a tunable $O(a^2)$ correction
$\tau_0$ to the gradient flow coupling, as suggested in Ref.~\cite{Cheng:2014jba}:
\begin{equation}
\label{eq:taucor}
 \gGF^2 = \frac{t^2}{\mathcal{N}} \langle E(t+\tau_0 a^2) \rangle = \frac{t^2}{\mathcal{N}} \langle E(t) \rangle + \frac{t^2}{\mathcal{N}} \langle \frac{\partial E(t)}{\partial t} \rangle \tau_0 a^2 + O(a^4).
\end{equation}
The detailed implementation of this procedure is described in sect.~\ref{sec:evolution}.

The running of the coupling is quantified using the finite lattice spacing step scaling function $\Sigma(u,s,L/a)$
and its continuum limit $\sigma(u,s)$ introduced in Ref.~\cite{Luscher:1992an}.
It describes the change of the measured coupling when the linear size of the system is increased from $L$ to $sL$,
while keeping the bare coupling $g_0^2$ (and hence the lattice spacing) constant:
\begin{align}
 &\Sigma(u,s,L/a) = \left. \gGF^2(g_0^2,sL/a) \right |_{\gGF^2(g_0^2,L/a)=u}
 \label{eq:stepscaling}\\
 &\sigma(u,s) = \lim_{a/L\rightarrow 0} \Sigma(u,s,L/a),
\end{align}
where $u$ denotes $\gGF^2$ as measured from the smaller volume.
In this work we choose $s=2$.
The step scaling function is related to the $\beta$-function by
\begin{align}
  -2\ln(s) = \int_u^{\sigma(u,s)} \frac{dx}{\sqrt x \beta(\sqrt x)}.
\end{align}
Close to the fixed point, where the running is slow and $|\beta|$ small, we can approximate the $\beta$-function by
\begin{align}
  \beta(g) \approx \bar\beta(g) =
 \frac{g}{2\ln(s)} \left ( 1 - \frac{\sigma(g^2,s)}{g^2} \right ). \label{eq:beta*}
\end{align}
The estimating function $\bar\beta(g)$ is exact at a fixed point but deviates
from the actual $\beta$-function as $|g-g_\ast |$ becomes large.

\subsection{Measurement of the mass anomalous dimension}
\label{sec:mgamma}
We use two different methods to measure the mass anomalous dimension $\gamma= - d\ln m_q/d\ln\mu$,
the mass step scaling method allowed by Schrödinger functional boundary conditions
and spectral density method.
If the theory has an infrared fixed point,
the mass anomalous dimension at this point is independent of the scheme used.

\paragraph{Schrödinger functional step scaling method:}
We start by measuring the
pseudoscalar density renormalization constant on the lattice as~\cite{Sint:1998iq}
\begin{align}
\label{eq:zpzp}
Z_P(L) = \frac{\sqrt{2 f_1}}{f_P(L/2)},
\end{align}
where the pseudoscalar density correlation functions $f_P$ is defined in Eq.~\eqref{eq:fp} and
\begin{align}
f_1 &= \frac{-a^{12}}{2 L^6} \sum_{\bs u,\bs v, \bs y, \bs z}
  \ev{\bar \zeta'(\bs u)\gamma_5\lambda^a\zeta'(\bs v)\bar\zeta(\bs y)\gamma_5\lambda^a\zeta(\bs z)},
\end{align}%
where $\zeta'$, $\bar\zeta'$ are boundary fields defined at $x_0=L$.
The mass step scaling function is then defined as in Ref.~\cite{Capitani:1998mq}:
\begin{align}
 \Sigma_P(u,s,L/a) &=
    \left. \frac {Z_P(g_0,sL/a)}{Z_P(g_0,L/a)} \right |_{g^2(g_0,L/a)=u}
    \label{eq:Sigmap}\\
 \sigma_P(u,s) &= \lim_{a/L\rightarrow 0} \Sigma_P(u,s,L/a).
 \label{eq:Sigmapc}
\end{align}
Here we will choose $s=2$ as we did in Eq.~\eqref{eq:stepscaling}; indeed, the
same simulations provide configurations for both calculations.

The mass step scaling function is related to the anomalous dimension $\gamma$ by~\cite{DellaMorte:2005kg}
\begin{align}
 \sigma_P(u,s) = \left ( \frac{u}{\sigma(u,s)} \right ) ^{d_0/(2b_0)}
   \exp \left [-\int_{\sqrt u}^{\sqrt{\sigma(u,s)}} dx
   \left ( \frac{\gamma(x)}{\beta(x)} - \frac{d_0}{b_0 x} \right )   \right ],
   \label{massstep}
\end{align}
where $b_0=\beta_0/(16\pi^2)$ in terms of the one-loop coefficient $\beta_0=22/3-2 N_f/3$ of the beta function and
$d_0=3 C_2(F)/(8\pi^2)=9/(32\pi^2)$ is the corresponding one-loop coefficient for the anomalous dimension,
$\gamma_{\rm 1-loop} =d_0 g^2$.
Close to the fixed point the expression~\eqref{massstep} simplifies considerably:
if we denote the function estimating the anomalous dimension $\gamma(u)$ by
$\bar\gamma(u)$, we have
\begin{align}
\log \sigma_P(g^2,s) &\simeq -\bar\gamma(g^2)\int_\mu^{s\mu} \frac{d\mu'}{\mu'} = -\bar\gamma(g^2)\log s , \\
 &\Rightarrow \bar\gamma(g^2) = -\frac{\log \sigma_P(g^2,s)}{\log s }.
 \label{gammabar}
\end{align}
The estimator $\bar\gamma(g^2)$ is exact at a fixed
point $g^2 = g_\ast^2$, where $\beta(g^2)$ vanishes, and deviates from the actual anomalous dimension when $\beta$ is large. We denote
the anomalous exponent at the fixed point with
$\gamma_\ast = \bar\gamma(g_\ast^2) = \gamma(g_\ast^2)$.

\paragraph{Spectral density method:}
The scaling of the spectral density of the massless Dirac operator is
governed by the anomalous dimension of the mass~\cite{DelDebbio:2010ze}.
The explicit calculation of the eigenvalue distribution is numerically costly,
but recent advances in applications of stochastic methods~\cite{Giusti:2008vb}
have made the mode number of the Dirac operator numerically accessible.
This quantity allows the determination of the mass
anomalous dimension~\cite{Patella:2011jr}.

The mode number of the Dirac operator is defined in terms of the
eigenvalue density $\rho(\lambda)$:
\begin{equation}\label{moden:nu1}
\nu(\Lambda) = 2\int_0^{\sqrt{\Lambda^2 - m^2}} \rho(\lambda)\dd \lambda.
\end{equation}
At a fixed point $g^2 = g_\ast^2$
it follows the power law scaling behavior
\begin{equation}\label{modenumber1}
  \nu(\Lambda) \simeq \nu_0(m) +  C\left[\Lambda^2 - m^2\right]^{2/(1+\gamma_\ast)}
\end{equation}
in some intermediate energy range between the infrared and the ultraviolet in the vicinity of the fixed point.
The fit parameters are $\gamma_\ast $, the mass anomalous dimension at the fixed point, the fermion mass $m$, and constants $\nu_0(m)$ and $C$.
The range where this power law behavior holds is not
known \textit{a priori}, and needs to be determined by trial and error.
In principle, a theory with an infrared fixed point will flow towards the scaling behavior~\eqref{modenumber1} in the infrared, no matter what the UV coupling is.
However, on a single finite lattice only a limited range of scales are accessible.
Thus, the lattice coupling should be tuned so that the coupling at the scale of $\mu \sim 1/L$ is as close to the fixed point as possible.

We calculate the mode number per unit volume in Eq.~\eqref{moden:nu1} by using
\begin{equation}\label{moden:nu2}
\nu(\Lambda) =\lim_{V\to \infty}  \frac{1}{V}\left< \textnormal{tr }\mathbb{P}_\Lambda \right>,
\end{equation}
where the operator $\mathbb{P}_\Lambda$ projects from the full eigenspace of $M = m^2 - \slashed{D}^2$ to the eigenspace of eigenvalues less than $\Lambda^2$.
The trace is calculated stochastically,
\begin{equation}\label{moden:projection}
\textnormal{tr } \mathbb{P}_\Lambda\simeq\frac{1}{N} \displaystyle \sum_{i=1}^N (\eta_i, \mathbb{P}_\Lambda \eta_i),
\end{equation}
where $\eta_i$ are $N$ pseudofermion fields.
This is described in detail in appendix~\ref{app:mgamma}.

Because the fermion mass~\eqref{eq:pcacm} is tuned to zero,
these two parameters $\nu_0(m)$ and $m^2$ in Eq.~\eqref{modenumber1} are expected to be small.
Indeed, in practice we observe the two constants to be negligible, and in our analysis we use a fit ansatz of the form
\begin{equation}\label{moden:moden2}
\nu(\Lambda) \simeq  C\Lambda^{4/(1+\gamma)}.
\end{equation}
We have checked that the error relative to the form including all four parameters, Eq.~\eqref{modenumber1}, is $\mathcal{O}(10^{-3})$.
The fit range was determined by varying the lower and the upper limit of the fit range
and observing the stability of the fit and the parameter values and their errors.

\section{Simulations and results}
\label{measurements}

We use hybrid Monte Carlo (HMC) simulation algorithm
with 2nd order Omelyan integrator~\cite{Omelyan:2003:SAI,Takaishi:2005tz}
and chronological initial values for the fermion matrix inversions~\cite{Brower:1995vx}.
The trajectory length is taken to be 1, and the step length is tuned to have an acceptance rate larger than 80\%.

We use lattices of volumes $(L/a)^4 = 6^4$, $8^4$, $10^4$, $12^4$, $16^4$, $20^4$, $24^4$ and $32^4$,
chosen to allow step scaling with scaling factor $s=2$ and the simulations were carried out with 18 values for $\beta_L=4/g_0^2$
ranging between 8 and 0.4.
In table~\ref{table:kappa}, corresponding to each of these values of $\beta_L$,
we show the critical value of the hopping parameter, $\kappa_c(\beta_L)$,
determined by requiring that the PCAC fermion mass~\eqref{eq:pcacm} vanishes at lattices of size $24^4$.
These values of $\kappa_c(\beta_L)$ is then used for all lattice volumes.
In general, we observe large finite size effects on $(L/a)^4=6^4$ lattices and hence these are not used in the
final analysis.

\begin{table}
%\TABLE{
%\centering
\begin{tabular}{ll}
\hline
 $\beta_L$  & $\kappa_c$ \\ \hline
\hline
8    & 0.125307435050069 \\   
6    & 0.125452134243701 \\  
5    & 0.125590630318978 \\  
4    & 0.125833726509734 \\  
3    & 0.126301695421514 \\  
2    & 0.127329165457485 \\  
1.7  & 0.127885967693622 \\
1.5  & 0.128375672766415 \\
1.3  & 0.129010604974215 \\
\hline
\end{tabular}
\hspace{0.5cm}
\begin{tabular}{ll}
\hline
 $\beta_L$  & $\kappa_c$  \\ \hline
\hline
1    & 0.130374869159002  \\
0.9  & 0.130990832298533  \\
0.8  & 0.131727494527597  \\
0.7  & 0.132608779236301  \\ 
0.6  & 0.133664962983886  \\ 
0.55 & 0.134267867684544  \\ 
0.5  & 0.134939416622759  \\ 
0.45 & 0.135670680413224  \\ 
0.4  & 0.136470043334909  \\ 
\hline
\end{tabular}
\caption{Values of the critical hopping parameter $\kappa_c$ used in the simulations at each $\beta_L=4/g_0^2$.}
\label{table:kappa}
%}
\end{table}

\begin{figure}
%\centering
\includegraphics[width=8.6cm]{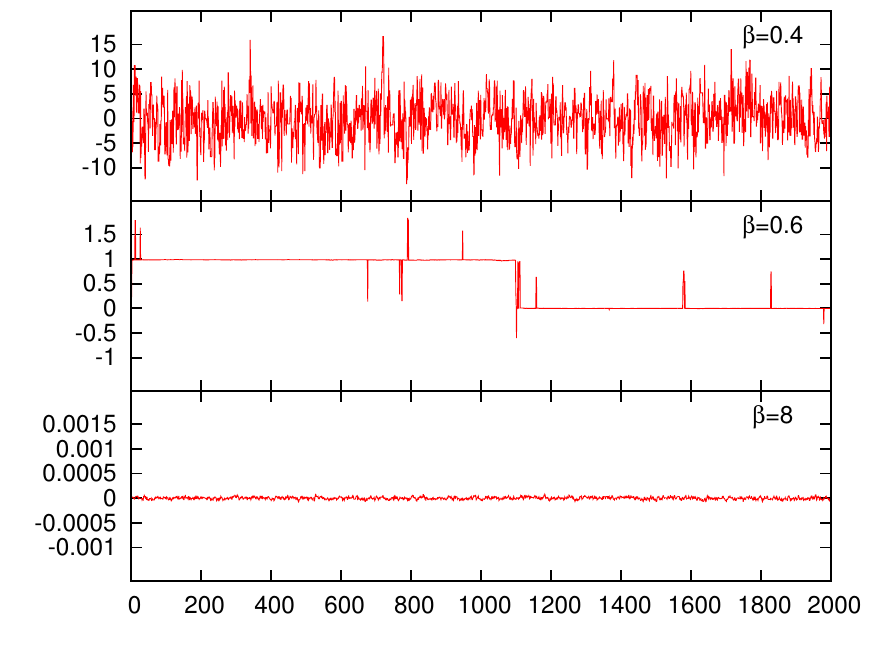} %previous figure: nonusedfigures/topofreeze.eps
\caption{Topological number at different values of $\beta_L$
         for the first 2000 trajectories in single configuration measured at $c_t=0.25$ and $L=32$.
		 The $\beta_L=0.6$ case shown in the middle is an example of the topology freezing.
		}
\label{fig:freezing}
\end{figure}%

The gradient flow method also allows us to measure the topological instanton number
without further computational costs from the cooled gauge fields
\begin{align}
Q(t) &= \frac{1}{32\pi^2} \sum_x \epsilon_{\mu\nu\alpha\beta}{G}_{\mu\nu}^a(x;t) G_{\alpha\beta}^a(x;t)
\label{eq:topodens}
\end{align}
for a large enough flow time $t$.
We use this to monitor the evolution of the topological number during the course of the simulations.
If the lattice fermion action were perfectly chiral,
the instanton would correspond to exact zero modes of the Dirac operator
and hence the instanton number freezes if the fermion mass is zero.
The stout smeared Wilson-clover action we use is expected to preserve chiral properties better than non-smeared Wilson-clover action.
Indeed, we observe that the topology is frozen
to $Q=0$ sector at small lattice couplings ($\beta_L \gsim 0.6$).
On the other hand, at very strong couplings the instanton number freely fluctuates around zero.
This is illustrated in Fig.~\ref{fig:freezing}.

Problems arise at the intermediate couplings,
where $Q$ can remain frozen in one topological sector for extended periods of simulation time before jumping to another one.
This can lead to very long autocorrelation times~\cite{Luscher:2014kea}.
The freezing is strongest at $\beta_L = 0.6$,
but we see mild freezing also on smaller values of $\beta_L$.
When the simulation is stuck in the non-zero topology sector
the measurements of e.g. the gradient flow coupling do not give sensible results.
Luckily, at $\beta_L=0.6$ where the metastability is strongest,
we generically observe that the system tunnels from a sector of nonzero $Q$ to the
sector of zero $Q$ and not vice versa.
Thus, in such a case we can interpret the nonzero $Q$ to be a thermalization effect,
and we can remove it by leaving out a sufficient number of trajectories from the beginning of the simulation.
This is shown in the center panel in Fig.~\ref{fig:freezing}.

After the thermalization 10 000-100 000 trajectories, for the smaller lattices, and
5000-30 000 trajectories, for the larger lattices, are left for the analysis.
The exact amount of generated trajectories are shown in table~\ref{table:meas}.

\begin{table}
\centering
\begin{tabular}{lllllllll}
\hline
$\beta_L$ & $L=8$ & $L=10$ & $L=12$ & $L=16$ & $L=20$ & $L=24$ & $L=32$ \\  \hline
\hline
8     & 108945  &  54663   &  70125   &  61344  &  74596  &  65708  &  62512   \\ 
6     & 46247   &  29528   &  31262   &  28606  &  21645  &  16319  &  6566    \\ 
5     & 31828   &  31662   &  29022   &  26761  &  9250   &  37593  &  9272    \\ 
4     & 31796   &  40720   &  46171   &  29973  &  38045  &  35794  &  32468   \\ 
3     & 117539  &  57963   &  70472   &  40970  &  49241  &  32703  &  3947    \\ 
2     & 31544   &  72688   &  29583   &  29181  &  28002  &  7084   &  13134   \\ 
1.7   & 111777  &  66772   &  81173   &  45333  &  43650  &  19052  &  8364    \\ 
1.5   & 85932   &  47137   &  52433   &  28258  &  23993  &  7204   &  9472    \\ 
1.3   & 75083   &  122133  &  106943  &  30709  &  42832  &  15103  &  13406   \\ 
1     & 227563  &  112698  &  45763   &  28957  &  41085  &  6011   &  7506    \\ 
0.9   & 101478  &  112544  &  67803   &  20638  &  42864  &  19174  &  9977    \\ 
0.8   & 53063   &  54667   &  29810   &  42741  &  26987  &  16587  &  14985   \\ 
0.7   & 156930  &  53515   &  53077   &  42967  &  27449  &  15991  &  27663   \\ 
0.6   & 72355   &  70660   &  67403   &  58410  &  27312  &  31122  &  16688   \\ 
0.55  & 81968   &  86878   &  76843   &  65762  &  52883  &  34169  &  11392   \\ 
0.5   & 80105   &  83670   &  67186   &  19639  &  23309  &  25643  &  16678   \\ 
0.45  & 78382   &  81749   &  72711   &  80815  &  50827  &  46446  &  13124   \\ 
0.4   & 75660   &  78431   &  68777   &  71500  &  60153  &  50014  &  16378   \\
\hline
\end{tabular}
\caption{Number of thermalized trajectories used for measurements performed for each lattice size.
         %These are numbers after cropping bad topology.
        }
\label{table:meas}
\end{table}

\subsection{Evolution of the coupling}
\label{sec:evolution}

\begin{figure}
%\centering
\includegraphics[width=8.6cm]{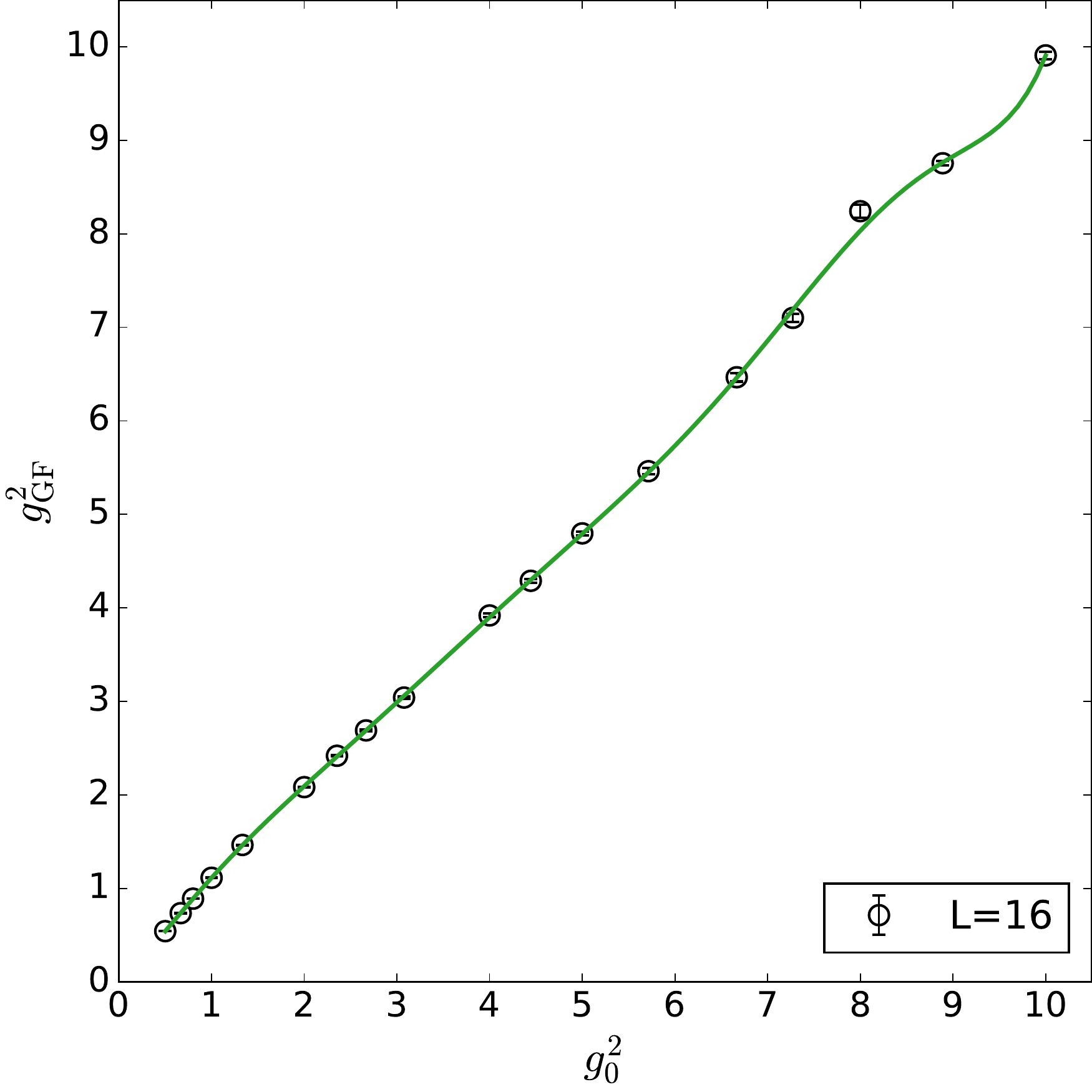} %betafit832.eps
\caption{Gradient flow coupling and the interpolating function~\eqref{eq:interpolation} for volume $(L/a)^4=16^4$}
\label{fig:betafit}
\end{figure}%

We measure the coupling using the gradient flow
method. As described in section~\ref{sec:couplingdefs}, we measure
the energy along gradient flow on each of the lattices shown on table~\ref{table:meas}.
A proper continuum extrapolation requires the step scaling function to be evaluated at constant coupling.
However, the simulations were done at a selected fixed set of bare couplings $\beta_L=4/g_0^2$,
which do not correspond to same $\gGF^2$-values when step scaling in Eq.~\eqref{eq:stepscaling}
is measured at different $L/a$.
Thus, it is necessary to interpolate the $g_0^2$-dependence of the actual measurements
of $\gGF^2(g_0^2,L/a)$ at each lattice size $L/a$,
and we use here a rational interpolating function~\cite{Karavirta:2011zg}:
\begin{align}
\label{eq:interpolation}
\gGF^2(g_0^2,L/a,t)=g_0^2\frac{1+\sum_{i=1}^{n} a_i g_0^{2i}}{1+\sum_{j=1}^{m} b_j g_0^{2j}}.
\end{align}

Because the small volumes quickly deviate from the tree level results at strong couplings,
relatively high order terms must be included in the fit.
However, at larger volumes there is a risk of overfitting,
especially since we observe some outlying points, that could indicate
underestimation of the statistical errors.
There is also no single choice of parameters $n,m$
that would give optimal $\chi^2/$d.o.f for all lattice sizes.
Therefore we find the set of parameters giving reasonable $\chi^2/$d.o.f and choose the most
probable one using the leave-one-out cross validation method.
This leads us to the parameters $n=7\,,m=1$, the result of which is demonstrated in figure~\ref{fig:betafit}.
The corresponding $\chi^2/$d.o.f for each volume are reported in table~\ref{table:betainterchi2}.
\begin{table}
\centering
\begin{tabular}{llllllll} \hline
$L/a$          & 8     & 10   & 12   & 16   & 20    & 24   & 32   \\ \hline
$\chi^2/d.o.f$ & 16.53 & 1.58 & 3.41 & 2.65 & 2.94  & 2.39 & 1.68 \\ \hline
\end{tabular}
\caption{The values of $\chi^2/\text{d.o.f}$ for each lattice size $L/a$}
\label{table:betainterchi2}
%}
\end{table}

Next we perform the continuum extrapolation to the step scaling function $\Sigma(u,2,L/a)$ defined in Eq.~\eqref{eq:stepscaling}.
Expecting the dominant discretization errors to be of order $a^2$,
we use a quadratic extrapolation function on lattices of size $L/a=10,12,16$:
\begin{align}
\Sigma(u,2,L/a)=\sigma(u,2)+c(u)(L/a)^{-2}
\label{eq:sigmaextr}
\end{align}%

\begin{figure}
%\centering
\includegraphics[width=8.6cm]{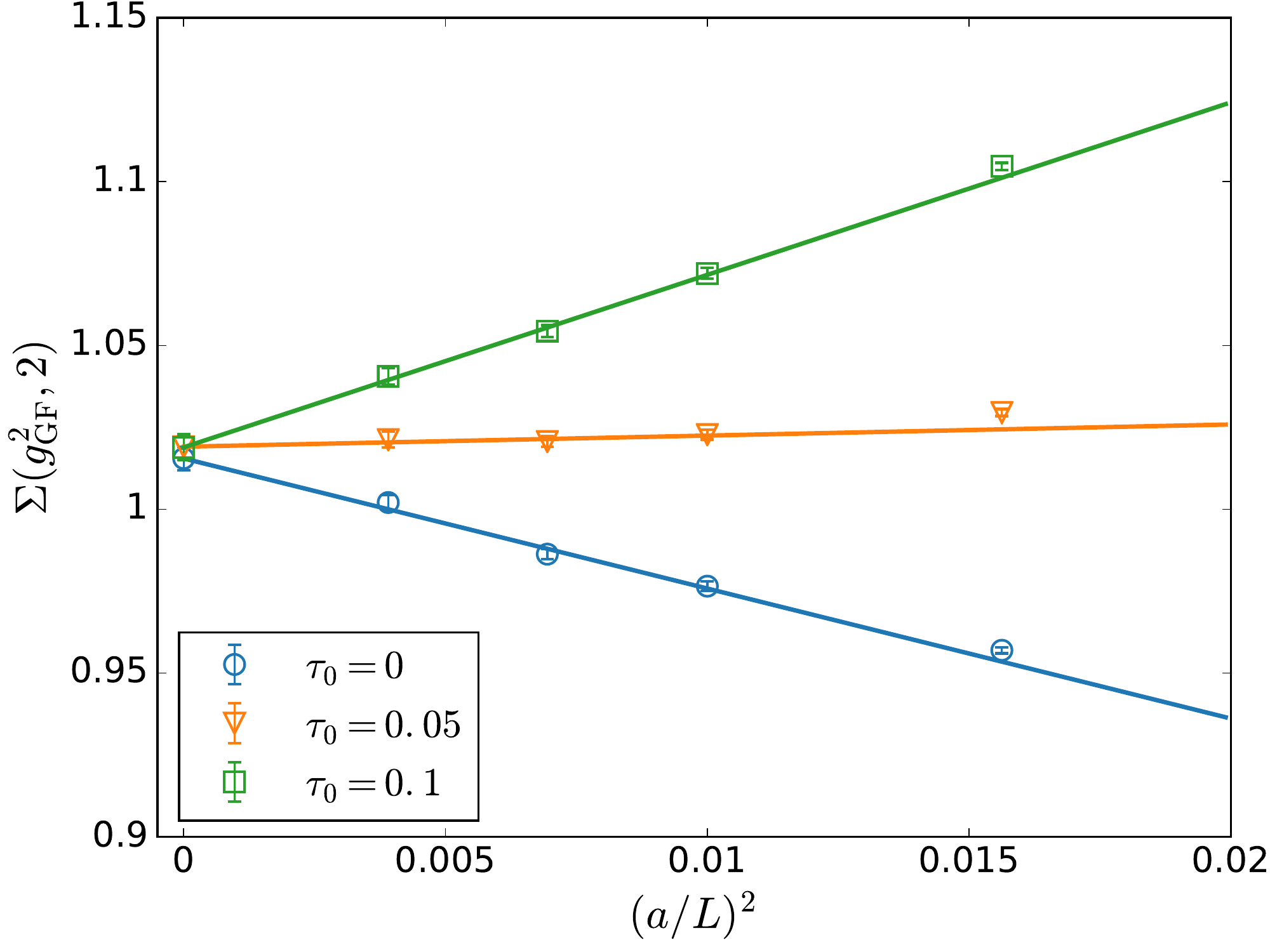}
\includegraphics[width=8.6cm]{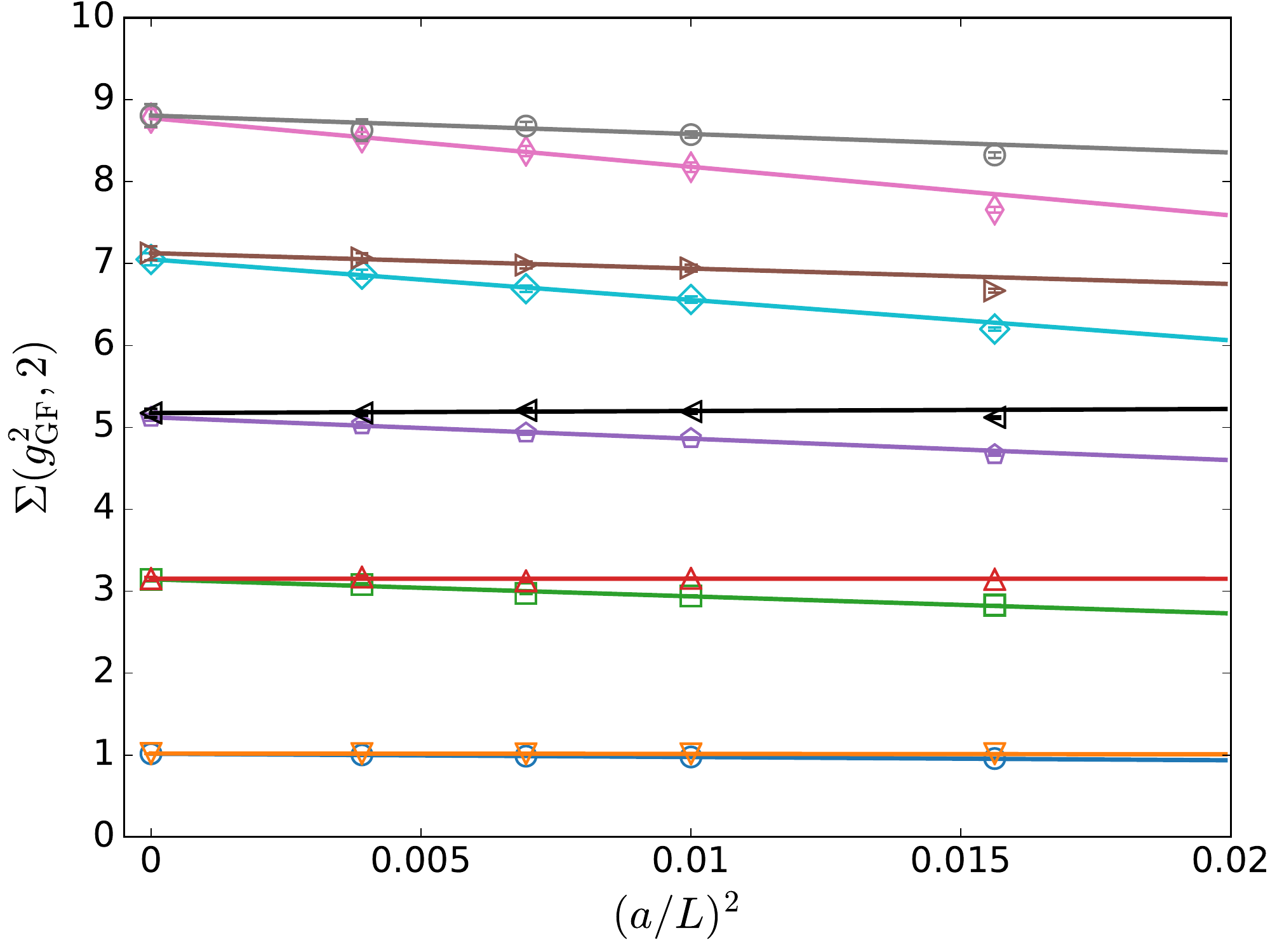}
\caption{Left: the $\tau_0$ correction process for $u=1$ with $c_t=0.4$. Right: continuum extrapolation of step scaling function
with and without $\tau_0$ correction at couplings $u=1,3,5,7,9$}
\label{fig:tau_fix}
\end{figure}%

In order to minimize the $\mathcal{O}(a^2)$ effects
we apply the $\tau_0$-correction to the gradient flow
trajectories as defined in Eq.~\eqref{eq:taucor}.  %eqref or cite to original paper?
In more detail, this is applied at the beginning of the analysis, i.e. for each new choice
of $\tau_0$ the coupling $\gGF$ and step scaling are recalculated.
A sample of the results is shown
in the figure~\ref{fig:tau_fix}.
It is clear that by suitably choosing $\tau_0$ most of the  $O(a^2)$ cutoff effects vanish.

As long as $|\tau_0|\ll t/a^2$,
the $\tau_0$-correction will have a relatively small effect in the continuum extrapolation~\cite{Hasenfratz:2014rna}.
The cutoff effects grow as a function of a coupling making the $\tau_0$-correction dependent on the coupling $\tau_0(\gGF^2)$.
In our case, at $c_t=0.4$ we have found that a good result can be obtained with the functional form
\begin{align}
\tau_0 = 0.06\log(1+\gGF^2)\,,
\label{eq:taufunc}
\end{align}
where the logarithmic form was chosen to regulate the behavior of $\tau_0$ at strong coupling.
In order to reach the final $\tau_0$ and $\gGF^2$ we calculate the correction iteratively starting from the bare coupling $g_0^2$.
For a consistent continuum limit the functional form must be of $\tau_0(\gGF^2)$ instead of, for example,
$\tau_0(g_0^2)$~\cite{Ramos:2015dla}.

\begin{figure}
%\centering
\includegraphics[width=8.6cm]{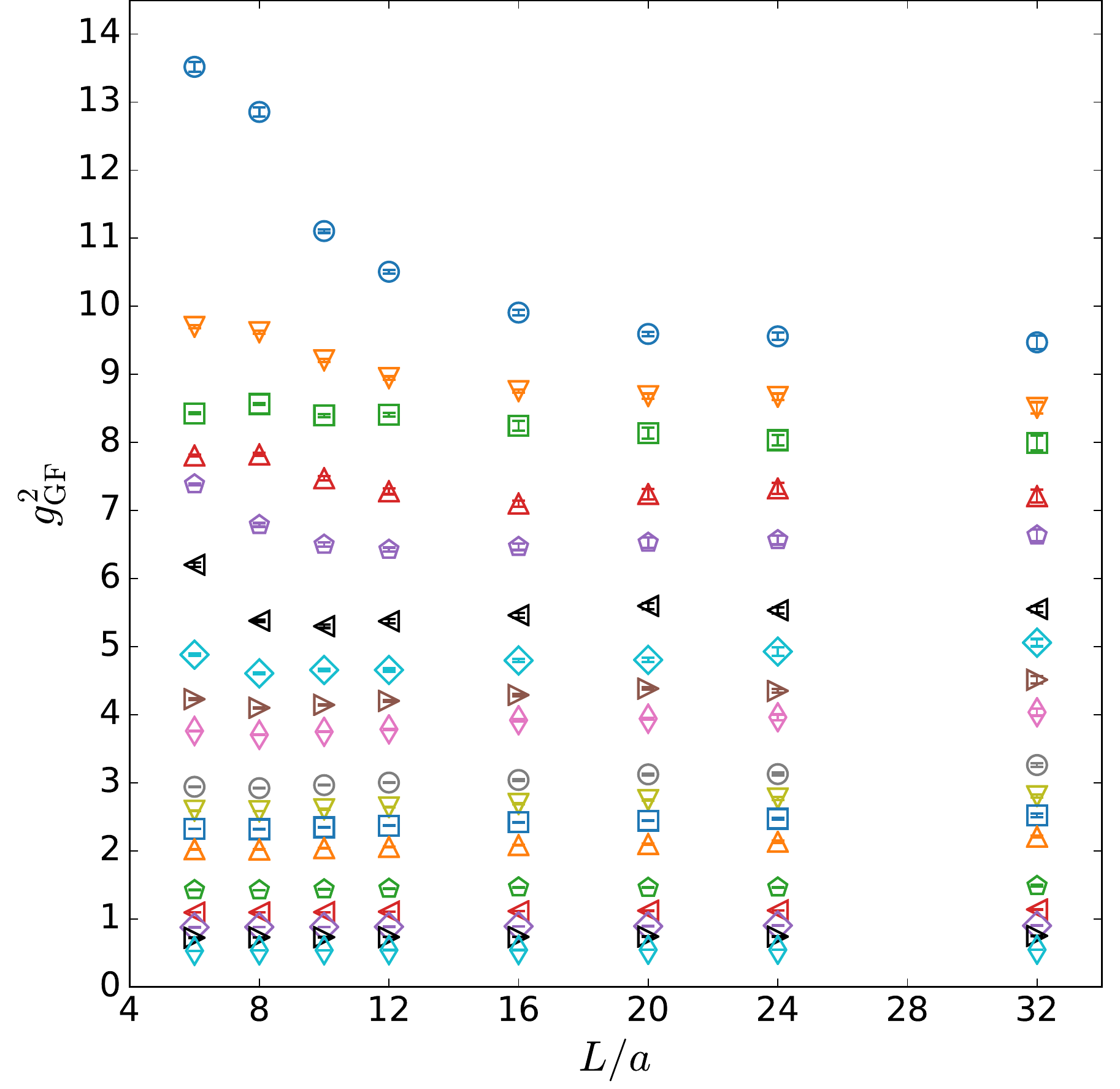} %previous figure: g2c04t06b.eps
\includegraphics[width=8.6cm]{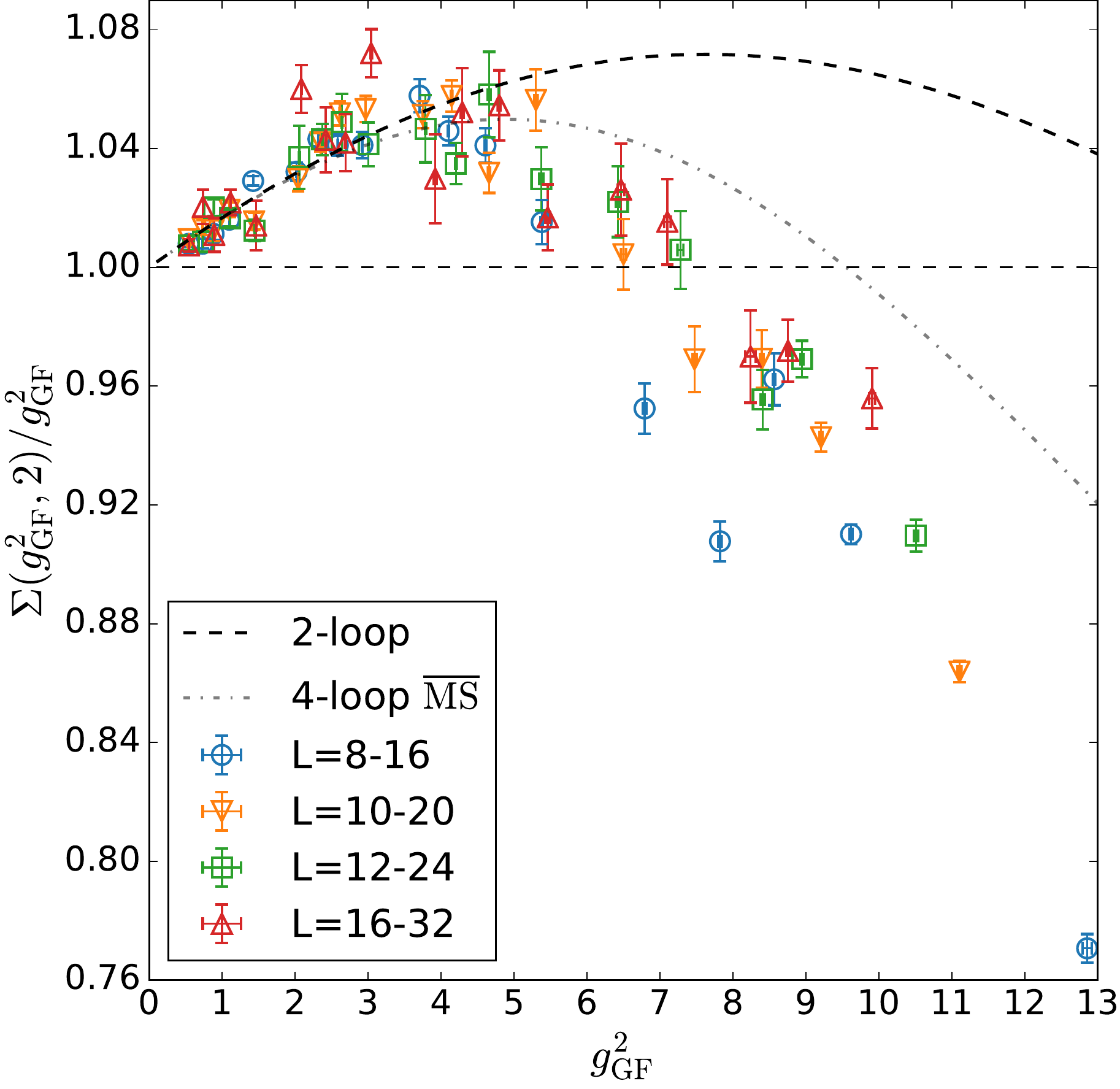} %previous figure: latssf.eps
\caption{Left: The measured values of $\gGF^2(g_0,L/a)$. Right: The lattice step scaling function.}
\label{fig:lattice_step_scaling_functions}
\end{figure}%

\begin{figure}
%\centering
\includegraphics[width=8.6cm]{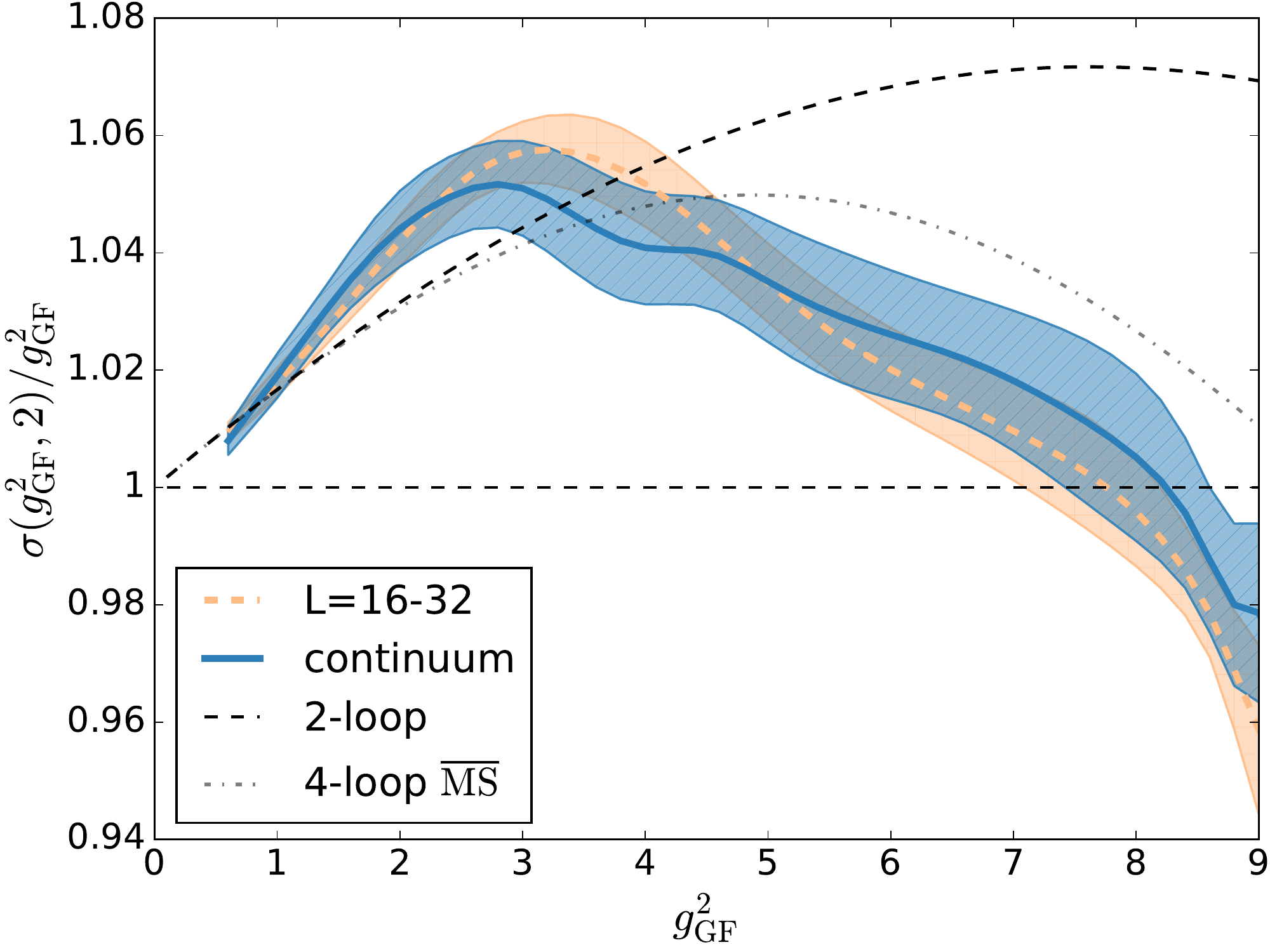}
\includegraphics[width=8.6cm]{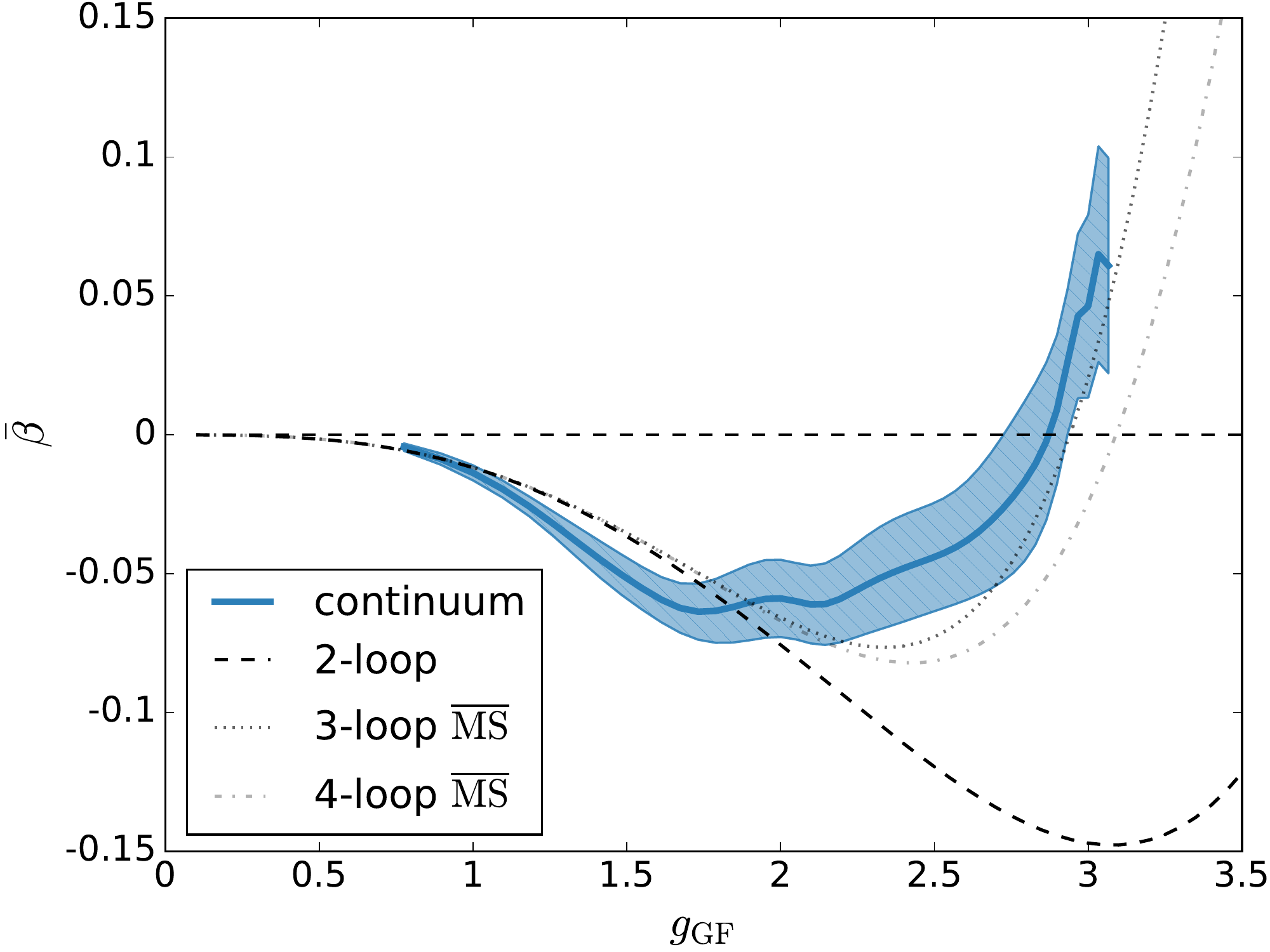}
\caption{Left: The scaled step scaling function $\sigma(\gGF^2,2)/\gGF^2$,
		 with continuum extrapolation done using the $10-20$, $12-24$ and $16-32$ volume pairs.
		 Right: The estimate of $\beta$-function.
     }
\label{fig:continuum_extrap1}
\end{figure}%

\begin{figure}
%\centering
\includegraphics[width=8.6cm]{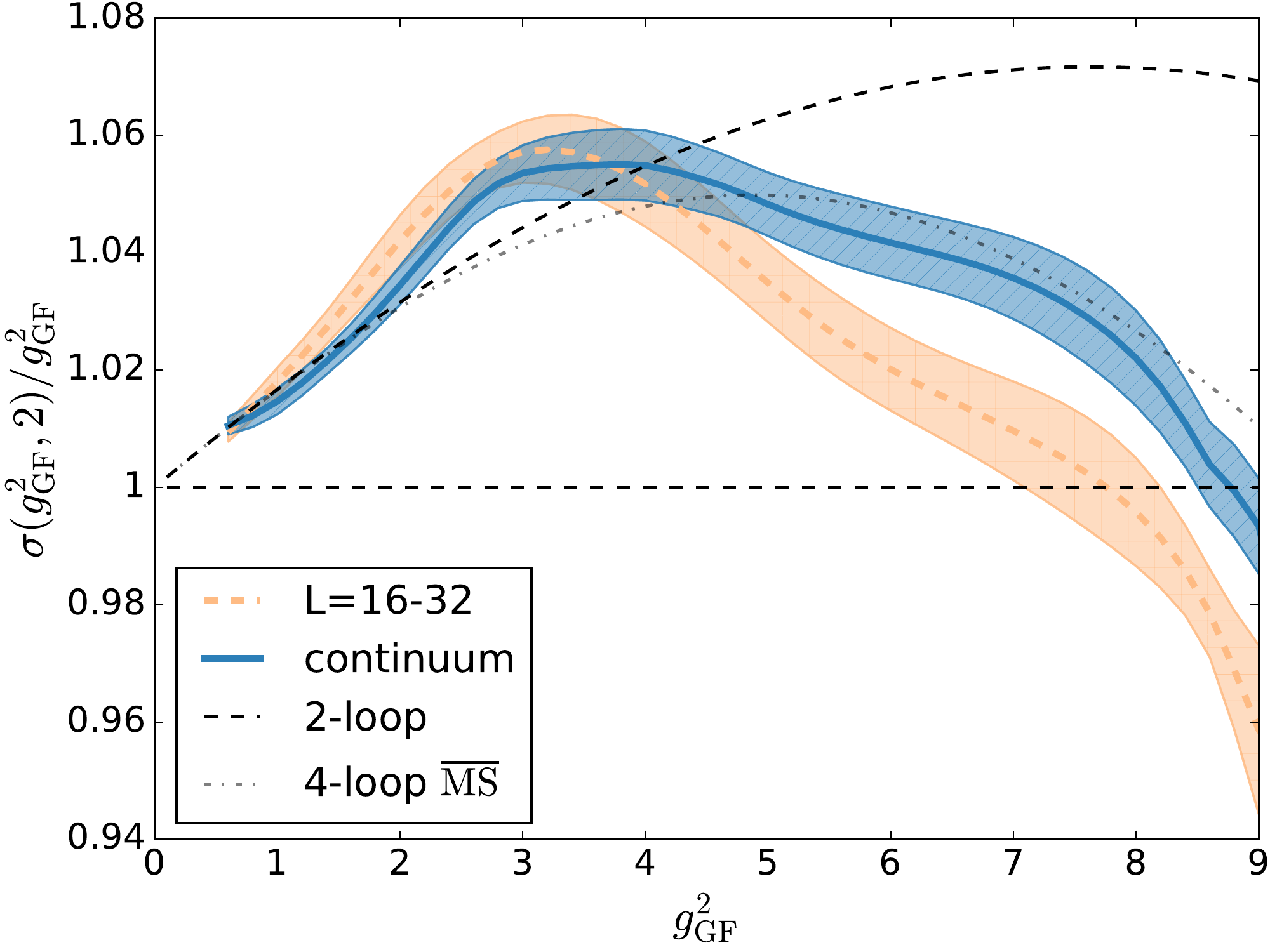}
\includegraphics[width=8.6cm]{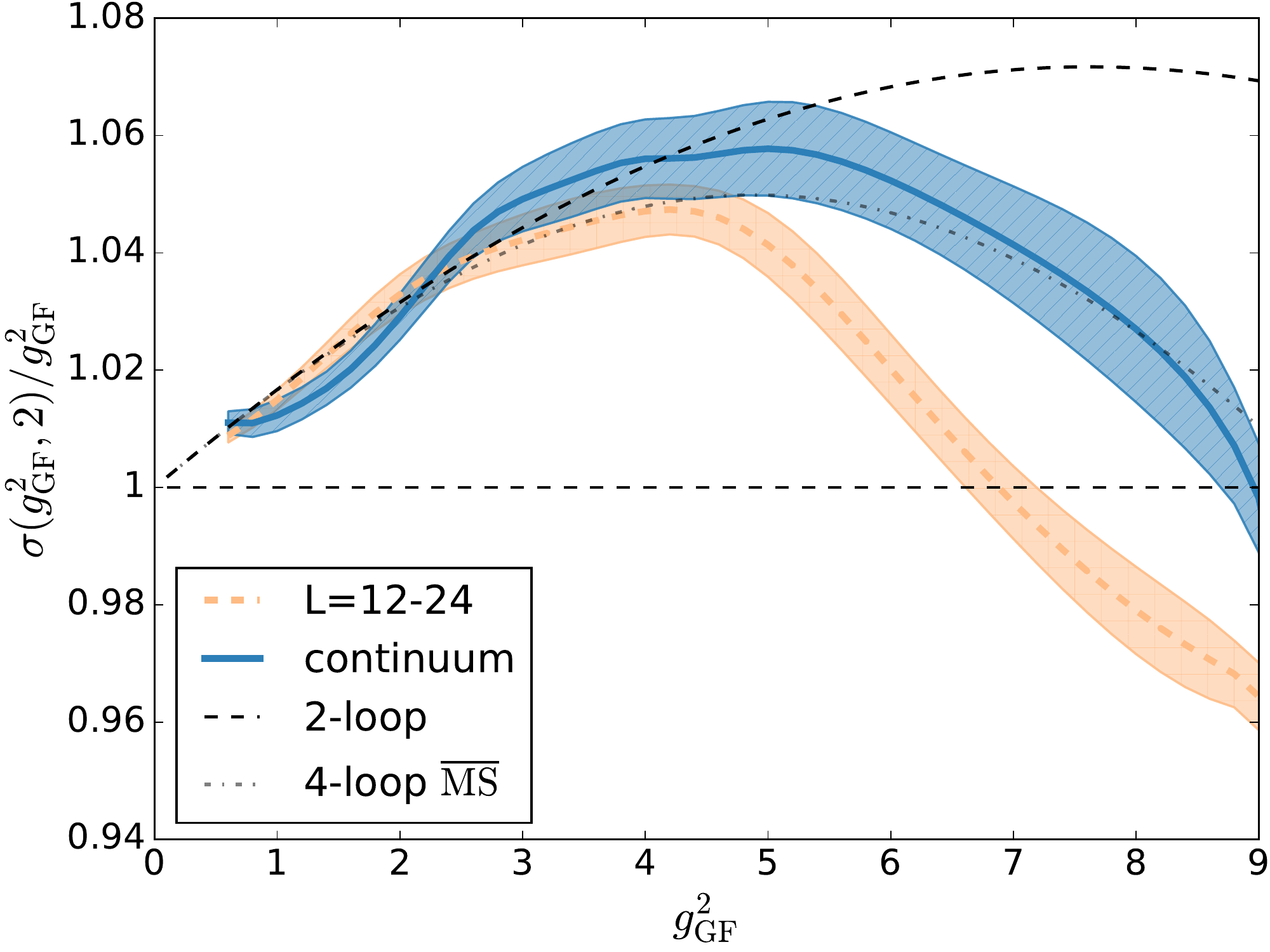}
\caption{The scaled step scaling function $\sigma(\gGF^2,2)/\gGF^2$,
		 with smallest volume pair $8-16$ included.
		 In the left panel the lattices up to $L/a=32$ are considered, while in the right
     panel the lattices only up to volume $L/a=24$.
		}
\label{fig:continuum_extrap2}
\end{figure}%

\begin{figure}
%\centering
\includegraphics[width=8.6cm]{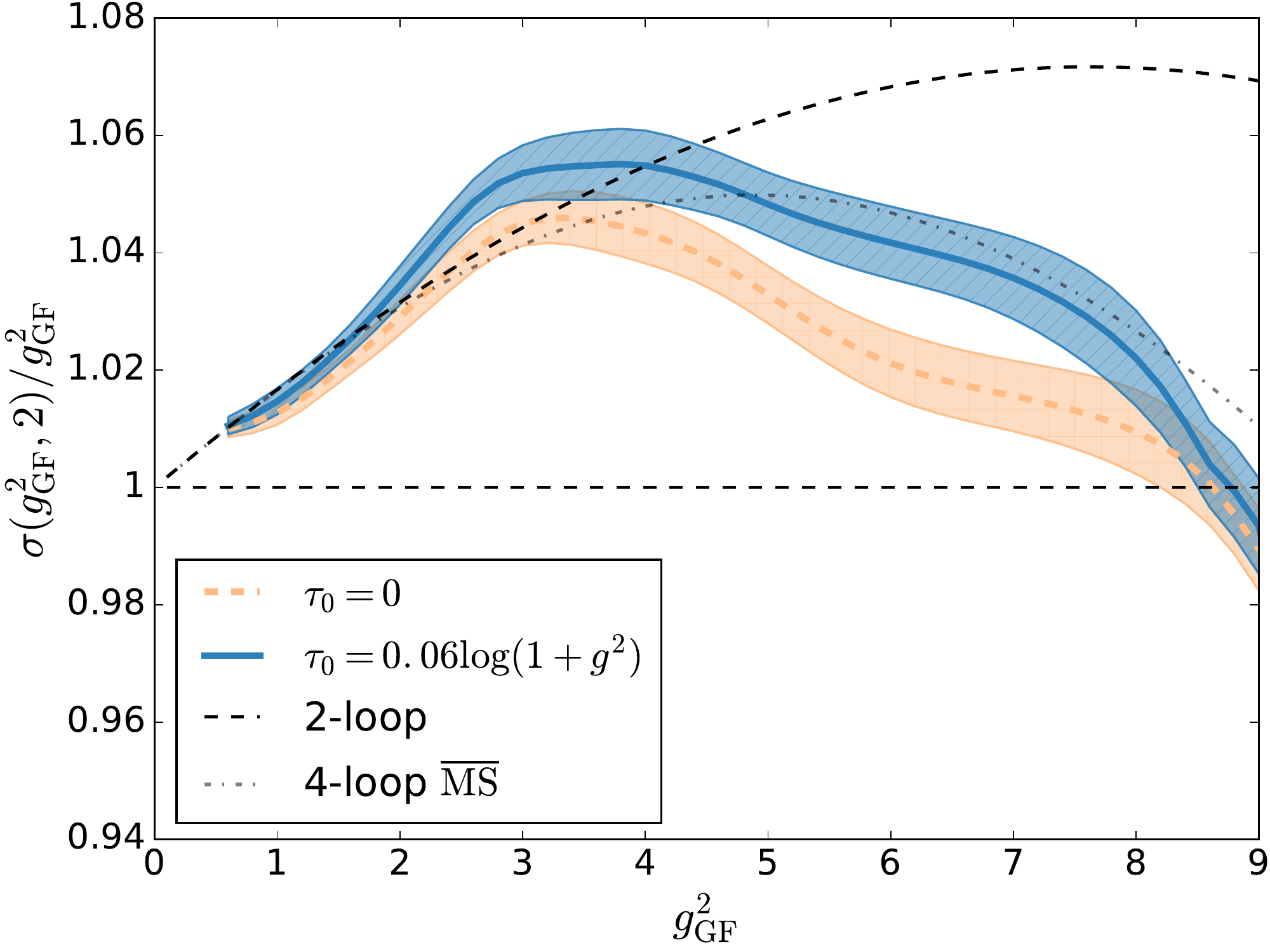}
\includegraphics[width=8.6cm]{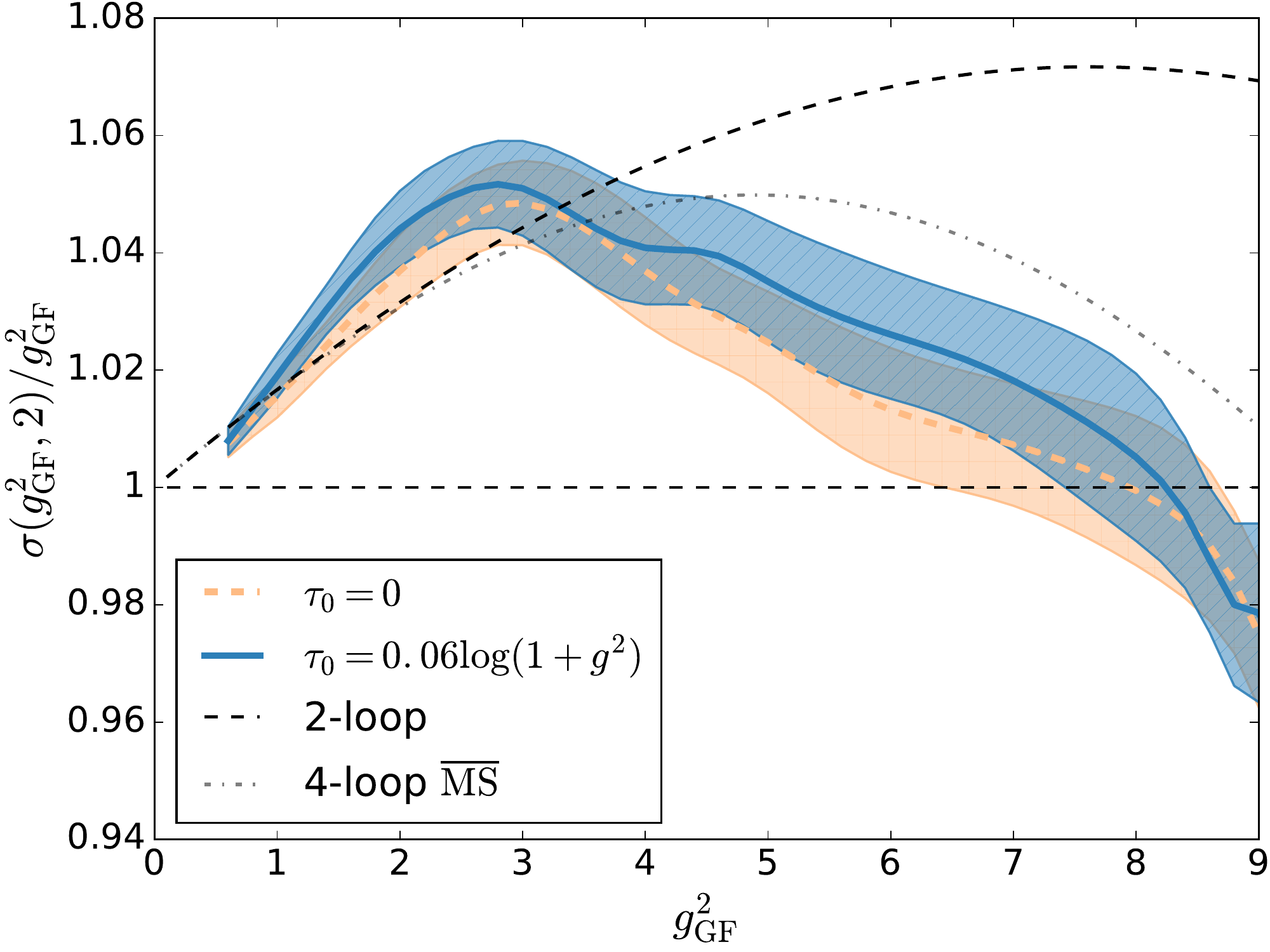}
\caption{Effect of $\tau_0$ correction to the continuum extrapolation.
		 The results in the left panel are obtained using volumes $8-32$ and in the right panel using
     volumes $10-32$.
		 As the correction was defined using only volumes $10-32$,
		 the continuum limit is affected by the correction when smaller volumes are included.
		}
\label{fig:continuum_extrap3}
\end{figure}%

\begin{figure}
%\centering
\includegraphics[width=8.6cm]{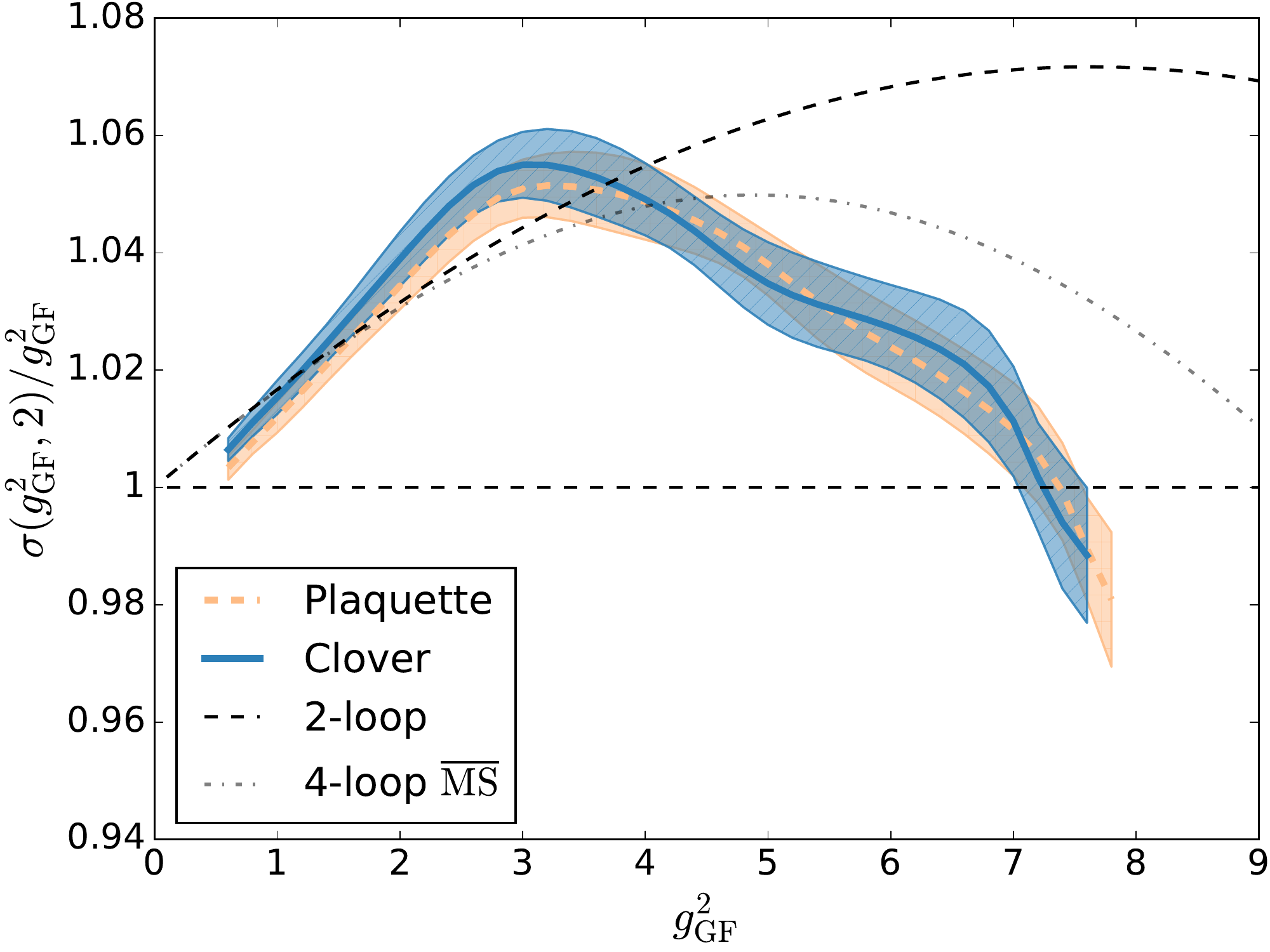}
\includegraphics[width=8.6cm]{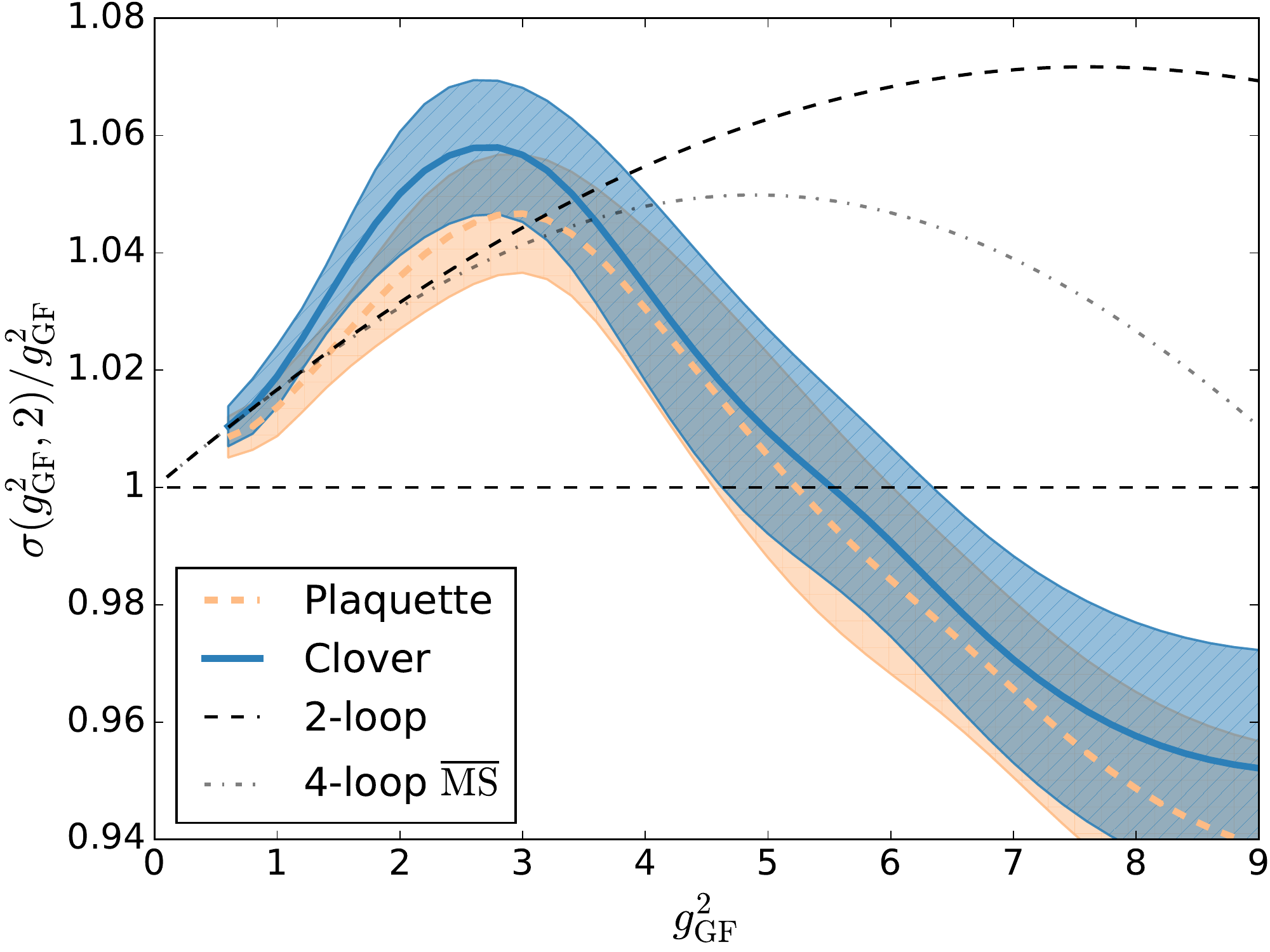}
\caption{Effect of the flow parameter $c_t$ to the continuum extrapolation:
		 $c_t=0.35$ (left), $c_t=0.5$ (right), with $\tau_0=0$ and volumes $10-32$.
		 We also show that the discretization of energy does not affect the continuum limit.
		}
\label{fig:continuum_extrap4}
\end{figure}%

The measured values of running coupling at $c_t=0.4$ with the $\tau_0$ correction
are given in tables~\ref{table:coupling} and~\ref{table:coupling2},
and shown in the left panel of the figure~\ref{fig:lattice_step_scaling_functions}.
We can see that the finite volume effects become substantial on smaller lattices at around $g^2 \equiv \gGF^2\approx 8$.
In the right panel of the figure we illustrate the scaled step scaling function $\Sigma(g^2,2,L/a)/g^2=g^2(g_0^2,2L/a)/g^2(g_0^2,L/a)$
at $L/a=8\,,10\,,12$ and $16$.
The running is compatible with the perturbation theory in the weak coupling region
following the universal 2-loop perturbative curve closely up to $g^2=4$,
but then deviating towards a possible fixed point around $g^2=8$.
While the 4-loop $\MSb$ result is scheme dependent and cannot be directly
compared with our result, it is nevertheless comforting to observe very comparable behavior.

We note that the small volume step scaling data at $L/a=8$ (we remind that $L/a=6$ is not included in the analysis at all)
shows unphysical structure, probably caused by finite volume effects.
This is especially evident in points near $g^2 = 7\ldots 10$, which appear to jump around erratically.
Thus, we will compare the extrapolations both with and without the $L/a=8$ case.

We present our final results in figures~\ref{fig:continuum_extrap1}--\ref{fig:continuum_extrap4}.
We check here the robustness of the result against changing the range of lattice volumes used in the extrapolation,
the use of $\tau_0$ correction, using the clover or plaquette definitions of $E$ in the gradient flow,
and the variation of the flow time parameter $c_t$.

In figure~\ref{fig:continuum_extrap1} we show our benchmark case,
the continuum limit of the step scaling function~\eqref{eq:sigmaextr}
using the $\tau_0$ correction from Eq.~\eqref{eq:taufunc} and
step scaling volume pairs $10$-$20$, $12$-$24$ and $16$-$32$ (thus excluding
the volume pair $8-16$).
We compare the continuum extrapolation with the largest volume step scaling function
which, in turn, can be compared to the uninterpolated step scaling presented in figure~\ref{fig:lattice_step_scaling_functions}.
The error bands shown include only the statistical errors from the measurements, interpolation and extrapolation.
The error propagation has been done by jackknife blocking throughout the whole analysis.
Thus, the variation between these two curves gives an estimate of the systematic errors in the extrapolation,
which seems to be well in control.
From the step scaling function $\sigma(\gGF^2)$ we can construct the
approximate beta function $\bar\beta(g)$, Eq.~\eqref{eq:beta*}.
This is shown in the right panel of figure~\ref{fig:continuum_extrap1}.

If we include the small volume $L/a=8$-$16$ step scaling in the continuum extrapolation
we obtain the result shown in the left panel of
figure~\ref{fig:continuum_extrap2}.
In turn, the result with the largest volume $L/a = 16$-$32$ excluded is shown in the right panel.
As we can observe, the extrapolations change, but the overall variation remains approximately within the 1-$\sigma$ error bands,
showing the robustness of the result.
For comparison, in the right panel of figure~\ref{fig:continuum_extrap2}
we also show the result from step scaling $L/a=12$-$24$ without the continuum limit.

In figure~\ref{fig:continuum_extrap3} we show the effect of the removal of the $\tau_0$ correction, Eq.~\eqref{eq:taucor}.
The $\tau_0$ correction has an effect on the continuum limit between $3\lsim\gGF^2\lsim 8$
when the step scaling $L/a=8$-$16$ is included;
however, the location of the fixed point stays at the $g_\ast^2=7.94\pm1.27$.
Without the inclusion of the small volume the effect of the $\tau_0$ correction remains within 1-$\sigma$ bands.

The $\tau_0$ correction helps to reach a reliable continuum limit, but, given perfect data,
it would not change the final result.
On the other hand, modifying the flow parameter $c_t$ corresponds to a different coupling constant scheme
and it will have an effect on the continuum limit.
The results presented above are obtained using $c_t=0.4$.
In figure~\ref{fig:continuum_extrap4} we show the continuum extrapolation using volumes $10$-$20$, $12$-$24$ and $16$-$32$
with $\tau_0=0$ and flow parameters $c_t=0.35$ and $c_t=0.5$.
We can observe that the overall structure of the step scaling function is preserved, but the value of the fixed point coupling
is changed to $g_\ast^2=7.23\pm0.19$ and $g_\ast^2=5.52\pm0.9$ for $c_t=0.35$ and $c_t=0.5$ respectively.
The errors increase rapidly as $c_t$ is increased above 0.5.
This kind of behavior of the step scaling function
has been observed before for different models~\cite{Lin:2015zpa,Hasenfratz:2015ssa}.

In figure~\ref{fig:continuum_extrap4} we also compare the plaquette and clover discretizations of
the energy observable $E(t)$, Eq.~\eqref{eq:edisc}.
Both discretizations are seen to give very similar result, and we present our results using the clover discretization.

Overall, we observe that the final extrapolation is remarkably robust against variation of the fit parameters.
In the $c_t=0.4$ scheme the fixed point coupling is located at $g_\ast^2=8.24(59)_{-1.64}^{+0.97}$,
where the first error is the statistical and the second includes the range of results from different choice of parameters.
The errors are dominated by the systematics of the extrapolation.

\subsection{Anomalous dimension of the mass}

\subsubsection{Mass step scaling}
\label{sec:mass_ss}
\begin{figure}
%\begin{center}
\includegraphics[width=8.6cm]{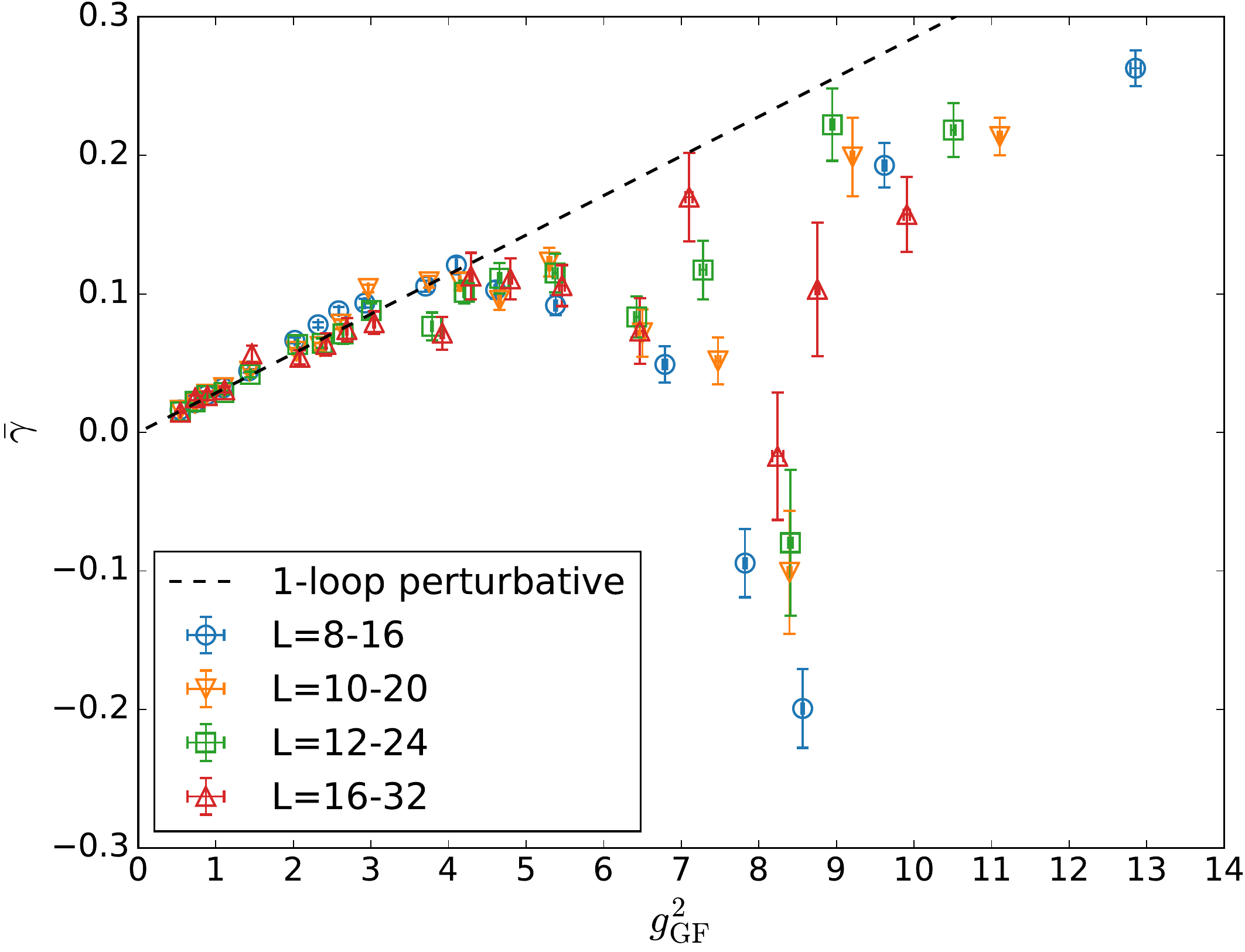}
\includegraphics[width=8.6cm]{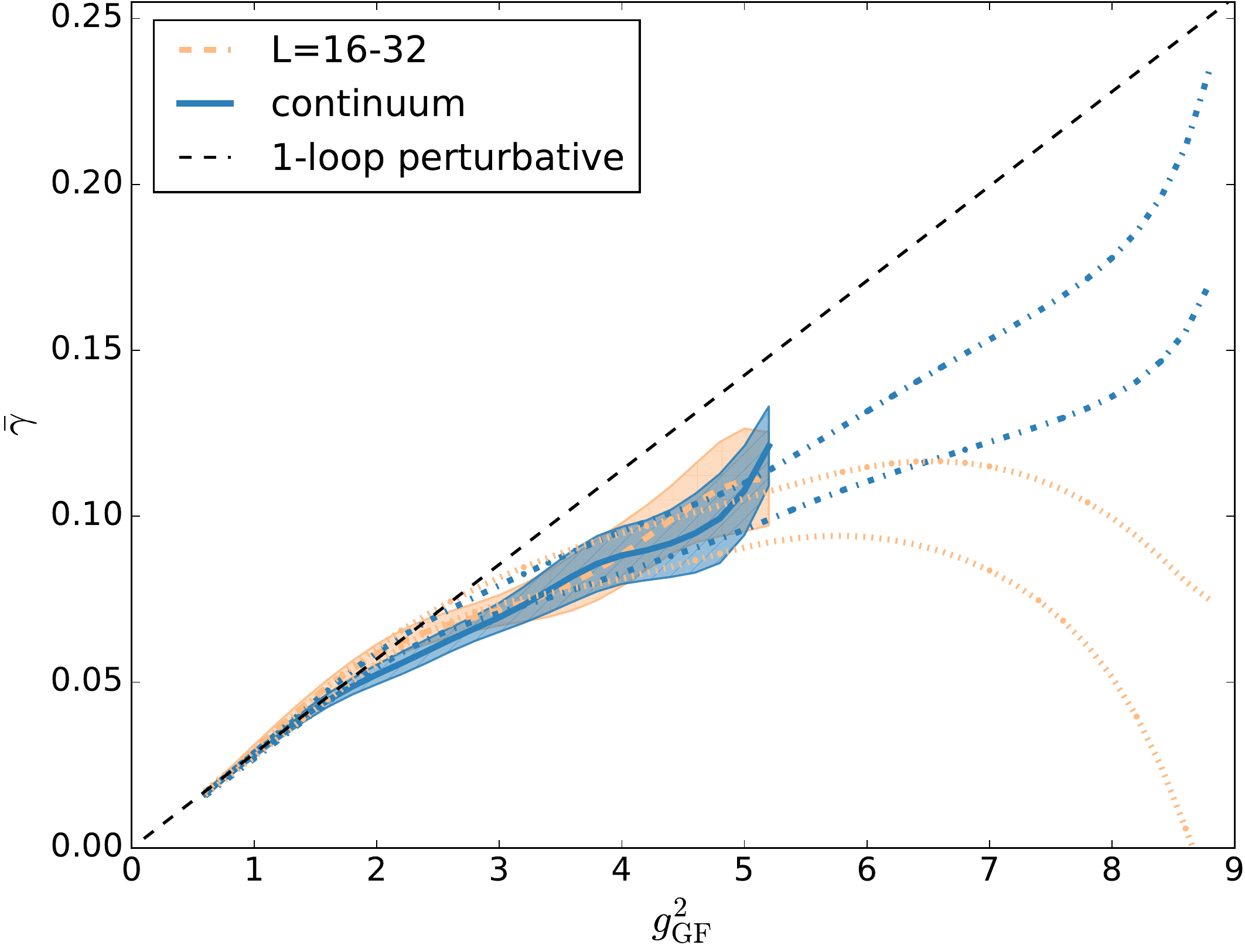}
%\end{center}
\caption{
  The estimate for the anomalous exponent $\bar\gamma$ from mass step scaling function (left) 
  and its continuum extrapolation (right). 
  The shaded bands show the result restricted to the region $\gGF^2 < 5.2$ where the continuum limit remains robust.  
  The dark shaded band corresponds to continuum limit using all volumes except $L/a=8$, 
  whereas the light shaded band is obtained using $L/a=16$ and $32$.  
  The empty dashed bands show the continuum limit up to $\gGF^2 < 8.8$.  
  In this case $\chi^2/$d.o.f of the fit Eq.~\eqref{eq:zpint} is unacceptably bad, 
  which is also evidenced by the large difference between the two bands.
}
\label{fig:stepgamma}
\end{figure}

The measurement of the anomalous dimension $\gamma$ using the mass step scaling method
described in section~\ref{sec:mgamma} is well established and has been applied
to many theories which may have an infrared fixed point, 
e.g. to SU(2) with fundamental fermions in Refs.~\cite{Bursa:2010xn,Karavirta:2011zg,Hayakawa:2013yfa}.
Our direct measurements of the estimate of the anomalous dimension $\bar\gamma(g^2)$, Eq.~\eqref{gammabar},
are shown in the left panel of figure~\ref{fig:stepgamma} at different
volumes, plotted against the measured coupling from the same pairs of volumes.  At small $\gGF^2$ the
measured estimate agrees well with the universal perturbative 1-loop curve.
However, at strong coupling, and especially as we approach the fixed point $\gGF^2 \sim 8$,
$\bar\gamma$ becomes dramatically smaller and we measure even negative values.
This behavior is caused by very strong finite size effects for this observable near the fixed point,
the magnitude of the negative peak is clearly reduced as the volumes grow.
Somewhat surprisingly, at even stronger coupling the measurements appear to stabilize again.
Nevertheless, these strong features in $\bar\gamma(g^2)$ make a controlled continuum limit questionable.

Despite these problems we attempt the continuum extrapolation in the right panel of figure~\ref{fig:stepgamma}.
The shaded bands show the continuum extrapolation Eq.~\eqref{eq:Sigmapc} and largest volume step scaling done
for couplings below the onset of strong finite size effects,
where the interpolation of the pseudoscalar density renormalization constant $Z_P$, Eq.~\eqref{eq:zpzp}:
\begin{equation}
\label{eq:zpint}
Z_P=1+\sum_{i=1}^n c_i g_0^{2i}\,,\quad n=5\,,
\end{equation}
gives $\chi^2$/\text{d.o.f}$\lsim 2$ for all used lattice sizes.
Again the volume pair $8-16$ is excluded.
In the same figure the dashed bands show how the continuum extrapolation and largest volume step scaling
would behave were the same interpolation done to all available bare couplings regardless of the goodness of the fit.
It is evident that the extrapolation is not under control near the fixed point coupling.

\subsubsection{Spectral density method}

Using the spectral density of the Dirac operator as described
in section~\ref{sec:mgamma} gives us a better controlled result than mass step scaling at strong coupling.
In this analysis
we use between 16 to 24 configurations of $L/a = 32$ lattices at eight values of the bare lattice coupling
$g_0^2 = 6/\beta = 0.75 \ldots 15$,  corresponding to measured gradient flow couplings $\gGF^2 \approx 0.75 \ldots 10$.
We calculate the mode number~\eqref{moden:nu1} for 100 values of $\Lambda^2$ ranging from $10^{-4}$ to $0.3$.

\begin{figure}
%\begin{center}
% These pictures were in eps, but changed to pdf's for arxiv submission
\includegraphics[width=8.6cm]{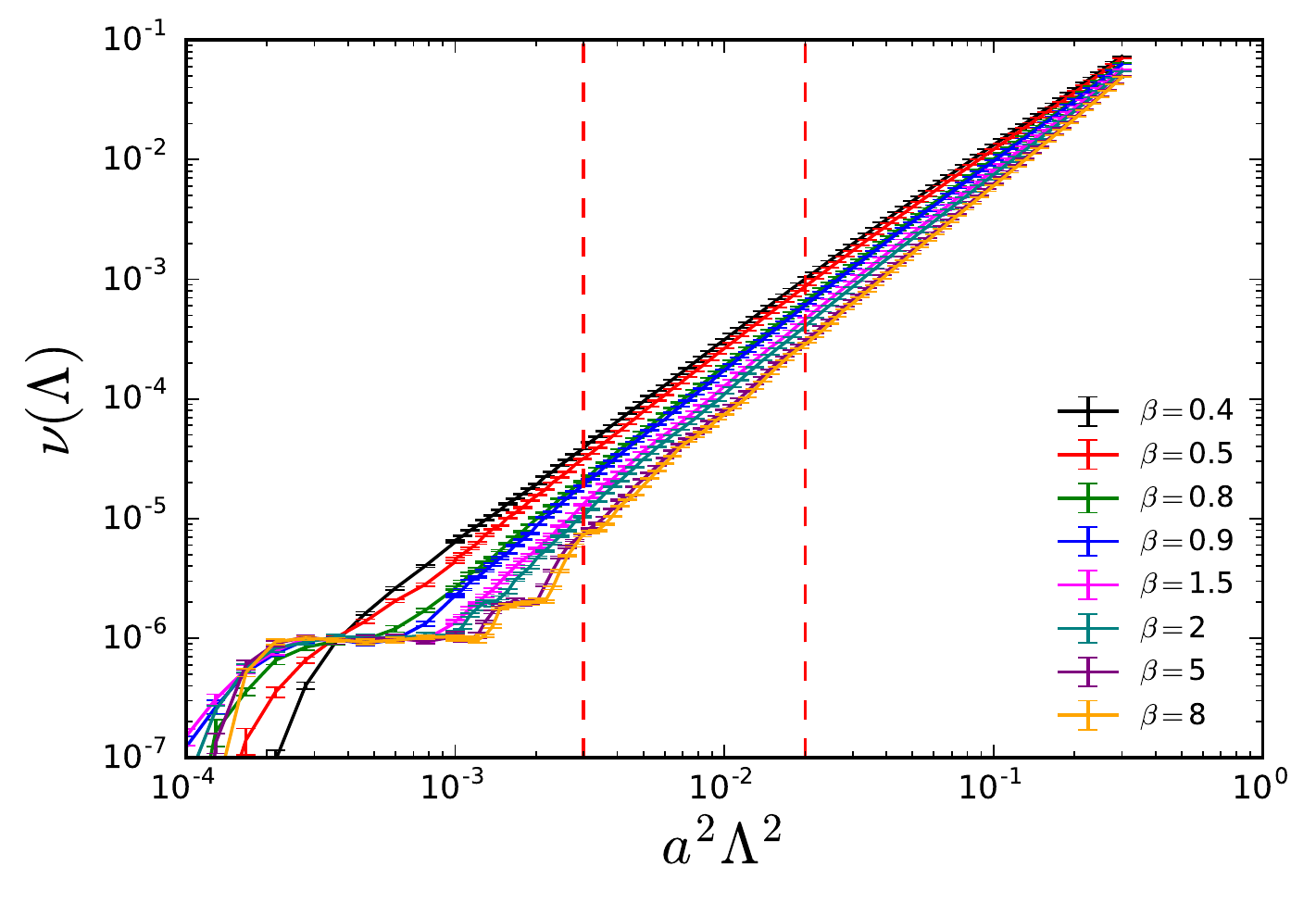}  
\includegraphics[width=8.6cm]{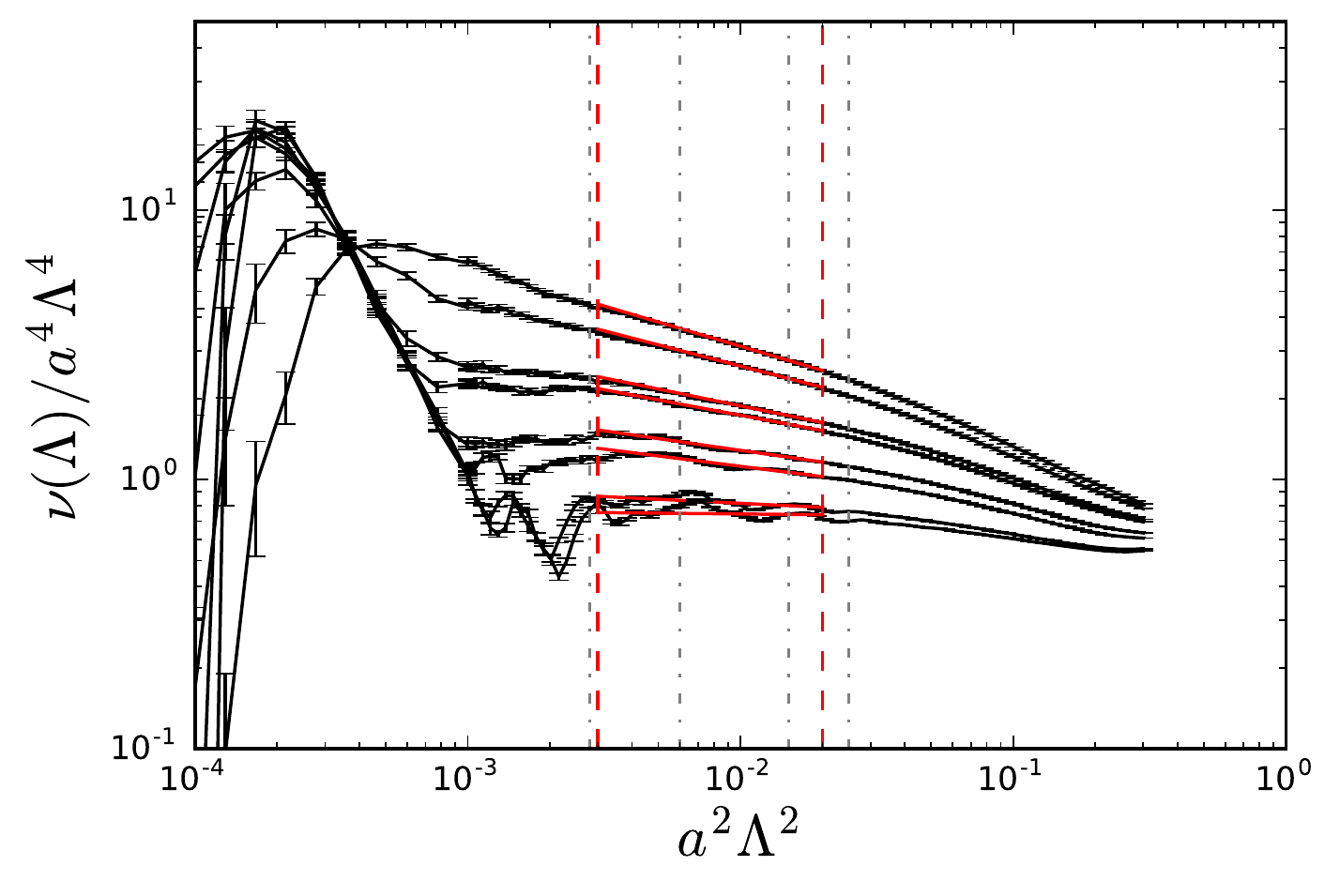}
\caption{Left: The mode number calculated for different gauge couplings on a $L/a = 32$ lattice. 
         Right: The mode number divided by $\Lambda^4$. 
		 The fit function and the fit range are indicated by solid and dashed red lines respectively. 
		 The curves are in the order of descending gauge coupling.}
\label{fig:modenumber_fit}
%\end{center}
\end{figure}

The raw mode number data is presented
in the left panel of figure~\ref{fig:modenumber_fit}.
The curves are in the order of descending gauge coupling.
At small couplings the behavior of the eigenvalues is close to the free fermion case,
and the lowest eigenvalues appear in discrete intervals,
manifested by the step-like structure of the mode number curve at small $\Lambda$.  This is
a finite volume effect, which becomes much milder
at couplings $\gGF^2 \ge 2.8$ where the interactions ``smear'' the eigenvalues more efficiently.

If the theory has an infrared fixed point, the mode number
behaves as $\nu \propto \Lambda^{4/(1+\gamma_\ast)}$ as $\Lambda \rightarrow 0$,
where $\gamma_\ast$ is the value of $\gamma$ at the fixed point.
In practice, finite volume effects limit the range of values of $\Lambda$ accessible on the lattice,
and in order to see the power law with the correct exponent the ensemble should be as close to the fixed point
as possible, i.e. the coupling measured from the ensemble should be close to the fixed point coupling.

\begin{figure}
%\begin{center}
\includegraphics[width=8.6cm]{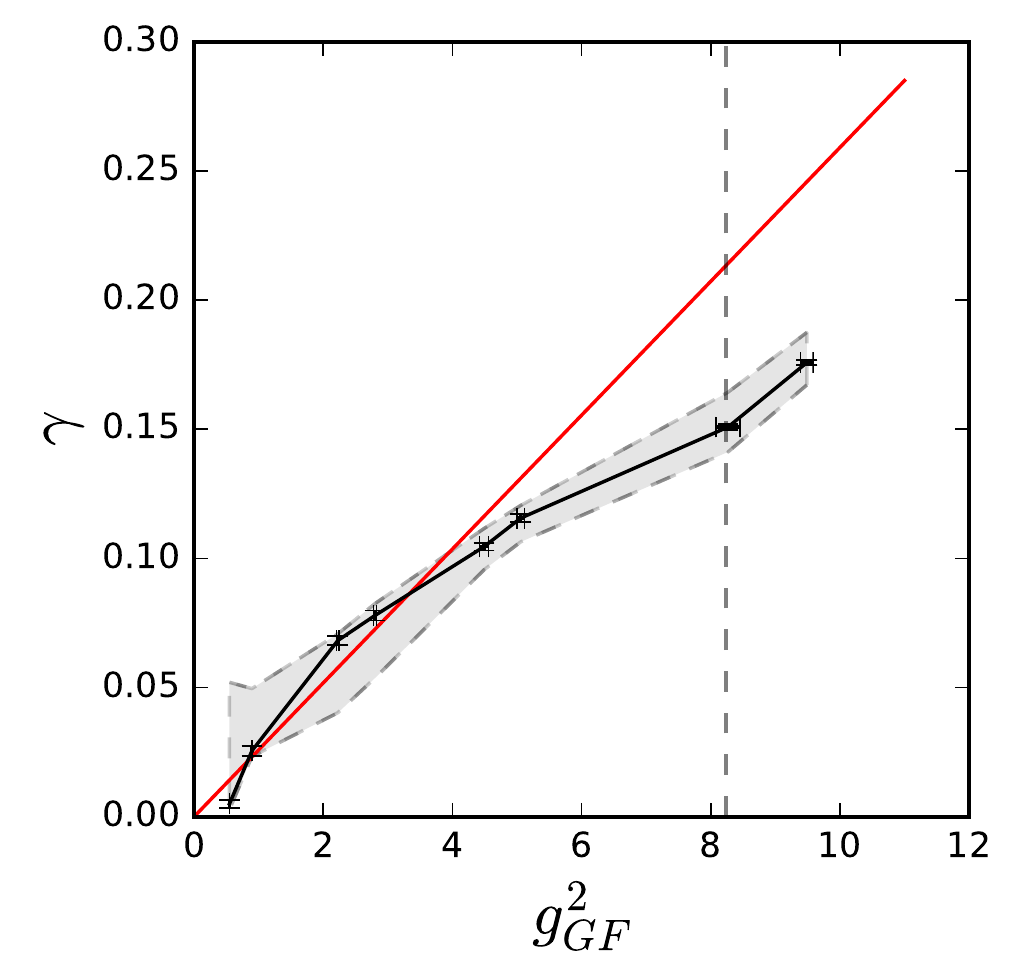}
\caption{The value of $\gamma$ obtained by fitting Eq.~\eqref{moden:moden2} 
         to the data in figure~\ref{fig:modenumber_fit} is shown with black points and the one loop perturbative result with a red line. 
		 The small error bars are statistical errors of the fits to the range $0.003 \le a^2\Lambda^2 \le 0.02$,
         and the shaded region is the error estimate obtained by varying the fit range as shown in figure~\ref{fig:modenumber_fit}.  
		 The dashed line indicates the location of the fixed point.}
\label{fig:ZPGFgamma}
%\end{center}
\end{figure}

In order to make the detailed behavior of the data visible,
we plot the mode number divided by the fourth power of the eigenvalue scale in the right panel of
figure~\ref{fig:modenumber_fit}.  In this case we expect the behavior
\begin{equation}
  \frac{\nu}{\Lambda^4} \propto \Lambda^{-4\gamma_\ast/(1+\gamma_\ast)}.
  \label{modefit}
\end{equation}
For the two strongest coupling ensembles, where we measure $\gGF^2$ close to $g_\ast^2 \approx 8$,
we observe a good power law behavior and
we can fit Eq.~\eqref{modefit} to the data
between $0.003 \le a^2\Lambda^2 \le 0.02$ with a reasonable $\chi^2/\text{d.o.f} \approx 1.5$.
The resulting exponents $\gamma$ are shown in figure~\ref{fig:ZPGFgamma},
with an estimated error range obtained by varying the fit range between the vertical lines shown in the figure,
which all give acceptable fits.
The statistical errors for a given fit are negligible in comparison with the uncertainty associated
with the variations of the fit range.

In order to obtain an estimate of the evolution of $\gamma(\gGF^2)$
we also fit the power law to ensembles of configurations at weaker couplings over the same range of 
$\Lambda$.
At weak couplings the fit quality becomes very poor due to the finite volume effects,
visible as a wave-like substructure on the right hand side of figure~\ref{fig:modenumber_fit}.
These features are a remnant of the discrete eigenvalue spectrum of the free theory.  The fitted value becomes 
very sensitive to the chosen fit range, increasing the estimated error on $\gamma(\gGF^2)$.
Nevertheless, the overall behavior of $\gamma$ as a function of $\gGF^2$ remains reasonable, 
as shown in figure~\ref{fig:ZPGFgamma}.  

At the estimated fixed point $\gGF^2 \approx 8.24 \pm 1.5$ we obtain the result $\gamma_\ast = 0.15\pm 0.02$,  
with the reservation that this result is obtained using only the largest $L/a=32$ lattices, i.e. a fixed lattice cutoff.  
The continuum limit is obtained by taking $L/a \rightarrow \infty$ limit while keeping $\gGF^2$ constant.  
Unfortunately, at volumes smaller than $L/a=32$ we do not obtain stable power law fits to the spectral density: 
the window of $a\Lambda$-values between the infrared finite size effects and the ultraviolet lattice spacing effects becomes too narrow.  
Reliable continuum limit would require simulations at significantly larger volumes, which would be prohibitively costly.

It is nevertheless interesting to observe 
that the above result is compatible with the continuum limit result obtained with the mass step scaling method, 
shown with dash-dotted lines in the right panel of figure~\ref{fig:stepgamma}.  
However, it should be remembered that the quality of the fit to Eq.~\eqref{eq:zpint} becomes very bad at $\gGF^2 \approx 8$, 
as discussed in section \ref{sec:mass_ss}.

\section{Conclusions}
\label{conclusions}

The determination of the lower boundary of the conformal window is a difficult non-perturbative problem,
with conflicting lattice results in the literature using both SU(2) and SU(3) gauge fields.
In this paper we studied SU(2) gauge theory with eight Dirac fermions in the fundamental representation of the
gauge group, using HEX-smeared Wilson-clover fermions and gradient flow method with Dirichlet boundary conditions.
Extrapolating our results to the continuum limit
we have established that the infrared properties of this theory are governed by a nontrivial fixed point
at $g_\ast^2 =8.24(59)_{-1.64}^{+0.97}$.
The result remains robust when different continuum extrapolations of the step scaling function are used.

We have also determined the mass anomalous dimension of the quark mass operator using two methods:
the Schr\"odinger functional mass step scaling function and the spectral density of the Dirac operator.
The mass step scaling is seen to become unreliable at strong coupling, whereas the spectral density remains stable,
and we obtain the mass anomalous dimension at the fixed point $\gamma_\ast = 0.15\pm 0.02$,
albeit using only the largest volume and thus a proper continuum limit is still lacking.

In the literature, there exists only one previous study of SU(2) gauge theory
with $N_f=8$ fundamental fermions by Ohki~{\it et al.}~\cite{Ohki:2010sr},
with inconclusive results about the existence of the fixed point.
Our result in this paper constitutes the first reliable result about the existence of the fixed point in this theory.
At $N_f=10$ the existence of the fixed point has been shown previously~\cite{Karavirta:2011zg}.
At $N_f=6$ the situation has been inconclusive~\cite{Karavirta:2011zg,Bursa:2010xn,Hayakawa:2013yfa,Appelquist:2013pqa},
but recent preliminary results indicate the existence of fixed point~\cite{Leino:2016njf}.

\FloatBarrier% Added float barriers to keep pics consentrated at the end, but before bibliography, remove if needed

\acknowledgments
This work is supported by the Academy of Finland
grants 114371 and 267842.
V.L and J.M.S. are supported by the Jenny and Antti Wihuri foundation,
and T.R. by the Magnus Ehrnrooth foundation.
J.R. acknowledges Danish National Research Foundation DNRF:90 grant.
The simulations were performed at the Finnish IT
Center for Science (CSC), Espoo, Finland,
on the Fermi supercomputer at Cineca in Bologna, Italy,
and on the K computer at Riken AICS in Kobe, Japan.
Parts of the simulation program have been derived from
the MILC lattice simulation program~\cite{MILC}.

\FloatBarrier

\appendix
\section{Mass anomalous dimension}
\label{app:mgamma}
The operator $\mathbb{P}_\Lambda$ in Eq.~\eqref{moden:nu2} can be approximated by
\begin{equation}\label{moden:p}
\mathbb{P}_\Lambda \simeq h(\mathbb{X})^4
\end{equation}
with $h(x)$ defined as
\begin{equation}
h(x) = \frac{1}{2}[1-xP(x^2)].
\end{equation}
Here $P(x)$ is a polynomial of degree $n$ that minimises the error
\begin{equation}
\delta = \underset{\epsilon \leq x \leq 1}{\max}\left| 1-\sqrt{x}P(x)\right|,
\end{equation}
and $\mathbb{X}$ in Eq.~\eqref{moden:p} is
\begin{equation}
\mathbb{X} = 1 - \frac{2\Lambda^2_\ast}{M + \Lambda^2_\ast},
\end{equation}
where $\Lambda_\ast$ is related to $\Lambda$ of Eq.~\eqref{moden:nu1} by
\begin{equation}
\frac{\Lambda}{\Lambda_\ast} = \left(\frac{ 1-\sqrt{\epsilon} }{1+\sqrt{\epsilon}}\right)^{1/2} +
\displaystyle \int_{-\sqrt{\epsilon}}^{\sqrt{\epsilon}} dx \frac{1+x}{(1-x^2)^{3/2}}h(x)^4.
\end{equation}
In our simulations we used
\begin{equation}
n = 32, \epsilon = 0.01, \delta \simeq 7.63\times 10^{-4}
\end{equation}
which gives us a ratio of
\begin{equation}
\frac{\Lambda}{\Lambda_\ast}  \simeq 0.9624.
\end{equation}
For $N$ which appears in Eq.~\eqref{moden:projection} we use $N=3$, 
since this was the number of pseudofermion fields for which the results seemed to converge.

\FloatBarrier % keep tables at tables section
\section{Tables}
\begin{table}[h]
\centering
\begin{tabular}{lllllllll}
 \hline
 $\beta_L$ & $L/a=6$  &  $L/a=8$            &  $L/a=10$           &  $L/a=12$           \\ \hline
\hline
8     &  0.5323(3)    &  0.5393(2)   &  0.5406(3)   &  0.5423(3)   \\  
6     &  0.7236(6)    &  0.7303(5)   &  0.7316(8)   &  0.7343(8)   \\  
5     &  0.8766(4)    &  0.8810(7)   &  0.8815(9)   &  0.8847(10)  \\  
4     &  1.0944(4)    &  1.0966(12)  &  1.0994(11)  &  1.1064(13)  \\  
3     &  1.4251(10)   &  1.4237(7)   &  1.4362(15)  &  1.4453(13)  \\  
2     &  2.0227(15)   &  2.018(2)    &  2.039(2)    &  2.057(5)    \\  
1.7   &  2.3243(17)   &  2.3186(19)  &  2.345(3)    &  2.370(3)    \\  
1.5   &  2.590(2)     &  2.583(2)    &  2.611(4)    &  2.640(5)    \\  
1.3   &  2.940(3)     &  2.921(4)    &  2.965(2)    &  3.003(4)    \\  
1     &  3.761(6)     &  3.705(3)    &  3.749(5)    &  3.786(9)    \\  
0.9   &  4.229(11)    &  4.101(8)    &  4.144(9)    &  4.203(10)   \\  
0.8   &  4.881(18)    &  4.607(13)   &  4.657(15)   &  4.65(2)     \\  
0.7   &  6.20(3)      &  5.379(17)   &  5.29(2)     &  5.37(3)     \\  
0.6   &  7.386(14)    &  6.78(2)     &  6.49(3)     &  6.42(3)     \\  
0.55  &  7.808(12)    &  7.82(2)     &  7.47(3)     &  7.28(4)     \\  
0.5   &  8.428(12)    &  8.564(16)   &  8.39(2)     &  8.40(2)     \\  
0.45  &  9.69(2)      &  9.61(2)     &  9.20(2)     &  8.94(2)     \\  
0.4   &  13.5(7)      &  12.8(6)     &  11.1(2)     &  10.5(2)     \\  
\hline
\end{tabular}
\caption{The measured values of $\gGF^2$ with $\tau$ correction applied, at each $\beta$ for small lattices $L/a=6\dots12$}
\label{table:coupling}
\end{table}
\begin{table}
\centering
\begin{tabular}{lllllllll}
\hline
$\beta_L$ & $L/a=16$ & $L/a=20$           & $L/a=24$           &  $L/a=32$          \\ \hline
\hline
8     &  0.5435(5)    &  0.5457(5)   &  0.5463(7)   &  0.5475(10)  \\  
6     &  0.7361(10)   &  0.7414(16)  &  0.740(2)    &  0.751(3)    \\  
5     &  0.8909(14)   &  0.892(3)    &  0.902(2)    &  0.900(4)    \\  
4     &  1.114(2)     &  1.120(2)    &  1.124(2)    &  1.138(4)    \\  
3     &  1.465(2)     &  1.458(3)    &  1.463(5)    &  1.485(12)   \\  
2     &  2.083(5)     &  2.100(7)    &  2.134(16)   &  2.211(16)   \\  
1.7   &  2.418(5)     &  2.444(6)    &  2.472(11)   &  2.52(2)     \\  
1.5   &  2.690(8)     &  2.747(10)   &  2.76(2)     &  2.80(2)     \\  
1.3   &  3.041(11)    &  3.123(12)   &  3.12(2)     &  3.24(2)     \\  
1     &  3.91(2)      &  3.941(14)   &  3.96(4)     &  4.03(5)     \\  
0.9   &  4.289(19)    &  4.383(17)   &  4.35(2)     &  4.51(5)     \\  
0.8   &  4.79(2)      &  4.80(3)     &  4.92(6)     &  5.05(5)     \\  
0.7   &  5.46(3)      &  5.59(4)     &  5.53(4)     &  5.55(4)     \\  
0.6   &  6.46(4)      &  6.52(7)     &  6.56(6)     &  6.63(8)     \\  
0.55  &  7.10(4)      &  7.24(7)     &  7.32(8)     &  7.20(9)     \\  
0.5   &  8.24(7)      &  8.13(8)     &  8.05(7)     &  7.99(10)    \\  
0.45  &  8.75(2)      &  8.68(3)     &  8.67(4)     &  8.50(8)     \\  
0.4   &  9.90(3)      &  9.59(3)     &  9.55(5)     &  9.47(10)    \\  
\hline
\end{tabular}
\caption{The measured values of $\gGF^2$ with $\tau$ correction applied, at each $\beta$ for large lattices $L/a=16\dots32$}
\label{table:coupling2}
\end{table}

\FloatBarrier

\bibliography{su2_nf8}{}

%merlin.mbs apsrev4-1.bst 2010-07-25 4.21a (PWD, AO, DPC) hacked
%Control: key (0)
%Control: author (72) initials jnrlst
%Control: editor formatted (1) identically to author
%Control: production of article title (-1) disabled
%Control: page (0) single
%Control: year (1) truncated
%Control: production of eprint (0) enabled
\begin{thebibliography}{50}%
\makeatletter
\providecommand \@ifxundefined [1]{%
 \@ifx{#1\undefined}
}%
\providecommand \@ifnum [1]{%
 \ifnum #1\expandafter \@firstoftwo
 \else \expandafter \@secondoftwo
 \fi
}%
\providecommand \@ifx [1]{%
 \ifx #1\expandafter \@firstoftwo
 \else \expandafter \@secondoftwo
 \fi
}%
\providecommand \natexlab [1]{#1}%
\providecommand \enquote  [1]{``#1''}%
\providecommand \bibnamefont  [1]{#1}%
\providecommand \bibfnamefont [1]{#1}%
\providecommand \citenamefont [1]{#1}%
\providecommand \href@noop [0]{\@secondoftwo}%
\providecommand \href [0]{\begingroup \@sanitize@url \@href}%
\providecommand \@href[1]{\@@startlink{#1}\@@href}%
\providecommand \@@href[1]{\endgroup#1\@@endlink}%
\providecommand \@sanitize@url [0]{\catcode `\\12\catcode `\$12\catcode
  `\&12\catcode `\#12\catcode `\^12\catcode `\_12\catcode `\%12\relax}%
\providecommand \@@startlink[1]{}%
\providecommand \@@endlink[0]{}%
\providecommand \url  [0]{\begingroup\@sanitize@url \@url }%
\providecommand \@url [1]{\endgroup\@href {#1}{\urlprefix }}%
\providecommand \urlprefix  [0]{URL }%
\providecommand \Eprint [0]{\href }%
\providecommand \doibase [0]{http://dx.doi.org/}%
\providecommand \selectlanguage [0]{\@gobble}%
\providecommand \bibinfo  [0]{\@secondoftwo}%
\providecommand \bibfield  [0]{\@secondoftwo}%
\providecommand \translation [1]{[#1]}%
\providecommand \BibitemOpen [0]{}%
\providecommand \bibitemStop [0]{}%
\providecommand \bibitemNoStop [0]{.\EOS\space}%
\providecommand \EOS [0]{\spacefactor3000\relax}%
\providecommand \BibitemShut  [1]{\csname bibitem#1\endcsname}%
\let\auto@bib@innerbib\@empty
%</preamble>
\bibitem [{\citenamefont {Appelquist}\ \emph {et~al.}(1986)\citenamefont
  {Appelquist}, \citenamefont {Karabali},\ and\ \citenamefont
  {Wijewardhana}}]{Appelquist:1986an}%
  \BibitemOpen
  \bibfield  {author} {\bibinfo {author} {\bibfnamefont {T.~W.}\ \bibnamefont
  {Appelquist}}, \bibinfo {author} {\bibfnamefont {D.}~\bibnamefont
  {Karabali}}, \ and\ \bibinfo {author} {\bibfnamefont {L.~C.~R.}\ \bibnamefont
  {Wijewardhana}},\ }\href {\doibase 10.1103/PhysRevLett.57.957} {\bibfield
  {journal} {\bibinfo  {journal} {Phys. Rev. Lett.}\ }\textbf {\bibinfo
  {volume} {57}},\ \bibinfo {pages} {957} (\bibinfo {year} {1986})}\BibitemShut
  {NoStop}%
%%CITATION = PRLTA,57,957;%%
\bibitem [{\citenamefont {Sannino}\ and\ \citenamefont
  {Tuominen}(2005)}]{Sannino:2004qp}%
  \BibitemOpen
  \bibfield  {author} {\bibinfo {author} {\bibfnamefont {F.}~\bibnamefont
  {Sannino}}\ and\ \bibinfo {author} {\bibfnamefont {K.}~\bibnamefont
  {Tuominen}},\ }\href {\doibase 10.1103/PhysRevD.71.051901} {\bibfield
  {journal} {\bibinfo  {journal} {Phys. Rev.}\ }\textbf {\bibinfo {volume}
  {D71}},\ \bibinfo {pages} {051901} (\bibinfo {year} {2005})},\ \Eprint
  {http://arxiv.org/abs/hep-ph/0405209} {arXiv:hep-ph/0405209 [hep-ph]}
  \BibitemShut {NoStop}%
%%CITATION = HEP-PH/0405209;%%
\bibitem [{\citenamefont {Pica}(2016)}]{Pica:2017gcb}%
  \BibitemOpen
  \bibfield  {author} {\bibinfo {author} {\bibfnamefont {C.}~\bibnamefont
  {Pica}},\ }\bibfield  {booktitle} {\emph {\bibinfo {booktitle} {{Proceedings,
  34th International Symposium on Lattice Field Theory (Lattice 2016):
  Southampton, UK, July 24-30, 2016}}},\ }\href@noop {} {\bibfield  {journal}
  {\bibinfo  {journal} {PoS}\ }\textbf {\bibinfo {volume} {LATTICE2016}},\
  \bibinfo {pages} {015} (\bibinfo {year} {2016})},\ \Eprint
  {http://arxiv.org/abs/1701.07782} {arXiv:1701.07782 [hep-lat]} \BibitemShut
  {NoStop}%
%%CITATION = ARXIV:1701.07782;%%
\bibitem [{\citenamefont {Hill}\ and\ \citenamefont
  {Simmons}(2003)}]{Hill:2002ap}%
  \BibitemOpen
  \bibfield  {author} {\bibinfo {author} {\bibfnamefont {C.~T.}\ \bibnamefont
  {Hill}}\ and\ \bibinfo {author} {\bibfnamefont {E.~H.}\ \bibnamefont
  {Simmons}},\ }\href {\doibase 10.1016/S0370-1573(03)00140-6} {\bibfield
  {journal} {\bibinfo  {journal} {Phys. Rept.}\ }\textbf {\bibinfo {volume}
  {381}},\ \bibinfo {pages} {235} (\bibinfo {year} {2003})},\ \bibinfo {note}
  {[Erratum: Phys. Rept.390,553(2004)]},\ \Eprint
  {http://arxiv.org/abs/hep-ph/0203079} {arXiv:hep-ph/0203079 [hep-ph]}
  \BibitemShut {NoStop}%
%%CITATION = HEP-PH/0203079;%%
\bibitem [{\citenamefont {Sannino}(2008)}]{Sannino:2008ha}%
  \BibitemOpen
  \bibfield  {author} {\bibinfo {author} {\bibfnamefont {F.}~\bibnamefont
  {Sannino}},\ }\href@noop {} {\  (\bibinfo {year} {2008})},\ \Eprint
  {http://arxiv.org/abs/0804.0182} {arXiv:0804.0182 [hep-ph]} \BibitemShut
  {NoStop}%
%%CITATION = ARXIV:0804.0182;%%
\bibitem [{\citenamefont {Karavirta}\ \emph {et~al.}(2012)\citenamefont
  {Karavirta}, \citenamefont {Rantaharju}, \citenamefont {Rummukainen},\ and\
  \citenamefont {Tuominen}}]{Karavirta:2011zg}%
  \BibitemOpen
  \bibfield  {author} {\bibinfo {author} {\bibfnamefont {T.}~\bibnamefont
  {Karavirta}}, \bibinfo {author} {\bibfnamefont {J.}~\bibnamefont
  {Rantaharju}}, \bibinfo {author} {\bibfnamefont {K.}~\bibnamefont
  {Rummukainen}}, \ and\ \bibinfo {author} {\bibfnamefont {K.}~\bibnamefont
  {Tuominen}},\ }\href {\doibase 10.1007/JHEP05(2012)003} {\bibfield  {journal}
  {\bibinfo  {journal} {JHEP}\ }\textbf {\bibinfo {volume} {05}},\ \bibinfo
  {pages} {003} (\bibinfo {year} {2012})},\ \Eprint
  {http://arxiv.org/abs/1111.4104} {arXiv:1111.4104 [hep-lat]} \BibitemShut
  {NoStop}%
%%CITATION = ARXIV:1111.4104;%%
\bibitem [{\citenamefont {Hietanen}\ \emph {et~al.}(2014)\citenamefont
  {Hietanen}, \citenamefont {Lewis}, \citenamefont {Pica},\ and\ \citenamefont
  {Sannino}}]{Hietanen:2014xca}%
  \BibitemOpen
  \bibfield  {author} {\bibinfo {author} {\bibfnamefont {A.}~\bibnamefont
  {Hietanen}}, \bibinfo {author} {\bibfnamefont {R.}~\bibnamefont {Lewis}},
  \bibinfo {author} {\bibfnamefont {C.}~\bibnamefont {Pica}}, \ and\ \bibinfo
  {author} {\bibfnamefont {F.}~\bibnamefont {Sannino}},\ }\href {\doibase
  10.1007/JHEP07(2014)116} {\bibfield  {journal} {\bibinfo  {journal} {JHEP}\
  }\textbf {\bibinfo {volume} {07}},\ \bibinfo {pages} {116} (\bibinfo {year}
  {2014})},\ \Eprint {http://arxiv.org/abs/1404.2794} {arXiv:1404.2794
  [hep-lat]} \BibitemShut {NoStop}%
%%CITATION = ARXIV:1404.2794;%%
\bibitem [{\citenamefont {Lewis}\ \emph {et~al.}(2012)\citenamefont {Lewis},
  \citenamefont {Pica},\ and\ \citenamefont {Sannino}}]{Lewis:2011zb}%
  \BibitemOpen
  \bibfield  {author} {\bibinfo {author} {\bibfnamefont {R.}~\bibnamefont
  {Lewis}}, \bibinfo {author} {\bibfnamefont {C.}~\bibnamefont {Pica}}, \ and\
  \bibinfo {author} {\bibfnamefont {F.}~\bibnamefont {Sannino}},\ }\href
  {\doibase 10.1103/PhysRevD.85.014504} {\bibfield  {journal} {\bibinfo
  {journal} {Phys. Rev.}\ }\textbf {\bibinfo {volume} {D85}},\ \bibinfo {pages}
  {014504} (\bibinfo {year} {2012})},\ \Eprint {http://arxiv.org/abs/1109.3513}
  {arXiv:1109.3513 [hep-ph]} \BibitemShut {NoStop}%
%%CITATION = ARXIV:1109.3513;%%
\bibitem [{\citenamefont {Ohki}\ \emph {et~al.}(2010)\citenamefont {Ohki},
  \citenamefont {Aoyama}, \citenamefont {Itou}, \citenamefont {Kurachi},
  \citenamefont {Lin}, \citenamefont {Matsufuru}, \citenamefont {Onogi},
  \citenamefont {Shintani},\ and\ \citenamefont {Yamazaki}}]{Ohki:2010sr}%
  \BibitemOpen
  \bibfield  {author} {\bibinfo {author} {\bibfnamefont {H.}~\bibnamefont
  {Ohki}}, \bibinfo {author} {\bibfnamefont {T.}~\bibnamefont {Aoyama}},
  \bibinfo {author} {\bibfnamefont {E.}~\bibnamefont {Itou}}, \bibinfo {author}
  {\bibfnamefont {M.}~\bibnamefont {Kurachi}}, \bibinfo {author} {\bibfnamefont
  {C.~J.~D.}\ \bibnamefont {Lin}}, \bibinfo {author} {\bibfnamefont
  {H.}~\bibnamefont {Matsufuru}}, \bibinfo {author} {\bibfnamefont
  {T.}~\bibnamefont {Onogi}}, \bibinfo {author} {\bibfnamefont
  {E.}~\bibnamefont {Shintani}}, \ and\ \bibinfo {author} {\bibfnamefont
  {T.}~\bibnamefont {Yamazaki}},\ }\bibfield  {booktitle} {\emph {\bibinfo
  {booktitle} {{Proceedings, 28th International Symposium on Lattice field
  theory (Lattice 2010)}}},\ }\href@noop {} {\bibfield  {journal} {\bibinfo
  {journal} {PoS}\ }\textbf {\bibinfo {volume} {LATTICE2010}},\ \bibinfo
  {pages} {066} (\bibinfo {year} {2010})},\ \Eprint
  {http://arxiv.org/abs/1011.0373} {arXiv:1011.0373 [hep-lat]} \BibitemShut
  {NoStop}%
%%CITATION = ARXIV:1011.0373;%%
\bibitem [{\citenamefont {Bursa}\ \emph {et~al.}(2011)\citenamefont {Bursa},
  \citenamefont {Del~Debbio}, \citenamefont {Keegan}, \citenamefont {Pica},\
  and\ \citenamefont {Pickup}}]{Bursa:2010xn}%
  \BibitemOpen
  \bibfield  {author} {\bibinfo {author} {\bibfnamefont {F.}~\bibnamefont
  {Bursa}}, \bibinfo {author} {\bibfnamefont {L.}~\bibnamefont {Del~Debbio}},
  \bibinfo {author} {\bibfnamefont {L.}~\bibnamefont {Keegan}}, \bibinfo
  {author} {\bibfnamefont {C.}~\bibnamefont {Pica}}, \ and\ \bibinfo {author}
  {\bibfnamefont {T.}~\bibnamefont {Pickup}},\ }\href {\doibase
  10.1016/j.physletb.2010.12.050} {\bibfield  {journal} {\bibinfo  {journal}
  {Phys. Lett.}\ }\textbf {\bibinfo {volume} {B696}},\ \bibinfo {pages} {374}
  (\bibinfo {year} {2011})},\ \Eprint {http://arxiv.org/abs/1007.3067}
  {arXiv:1007.3067 [hep-ph]} \BibitemShut {NoStop}%
%%CITATION = ARXIV:1007.3067;%%
\bibitem [{\citenamefont {Hayakawa}\ \emph {et~al.}(2013)\citenamefont
  {Hayakawa}, \citenamefont {Ishikawa}, \citenamefont {Takeda},\ and\
  \citenamefont {Yamada}}]{Hayakawa:2013yfa}%
  \BibitemOpen
  \bibfield  {author} {\bibinfo {author} {\bibfnamefont {M.}~\bibnamefont
  {Hayakawa}}, \bibinfo {author} {\bibfnamefont {K.~I.}\ \bibnamefont
  {Ishikawa}}, \bibinfo {author} {\bibfnamefont {S.}~\bibnamefont {Takeda}}, \
  and\ \bibinfo {author} {\bibfnamefont {N.}~\bibnamefont {Yamada}},\ }\href
  {\doibase 10.1103/PhysRevD.88.094504} {\bibfield  {journal} {\bibinfo
  {journal} {Phys. Rev.}\ }\textbf {\bibinfo {volume} {D88}},\ \bibinfo {pages}
  {094504} (\bibinfo {year} {2013})},\ \Eprint {http://arxiv.org/abs/1307.6997}
  {arXiv:1307.6997 [hep-lat]} \BibitemShut {NoStop}%
%%CITATION = ARXIV:1307.6997;%%
\bibitem [{\citenamefont {Appelquist}\ \emph {et~al.}(2014)\citenamefont
  {Appelquist}, \citenamefont {Brower}, \citenamefont {Buchoff}, \citenamefont
  {Cheng}, \citenamefont {Fleming}, \citenamefont {Kiskis}, \citenamefont
  {Lin}, \citenamefont {Neil}, \citenamefont {Osborn}, \citenamefont {Rebbi}
  \emph {et~al.}}]{Appelquist:2013pqa}%
  \BibitemOpen
  \bibfield  {author} {\bibinfo {author} {\bibfnamefont {T.}~\bibnamefont
  {Appelquist}}, \bibinfo {author} {\bibfnamefont {R.}~\bibnamefont {Brower}},
  \bibinfo {author} {\bibfnamefont {M.}~\bibnamefont {Buchoff}}, \bibinfo
  {author} {\bibfnamefont {M.}~\bibnamefont {Cheng}}, \bibinfo {author}
  {\bibfnamefont {G.}~\bibnamefont {Fleming}}, \bibinfo {author} {\bibfnamefont
  {J.}~\bibnamefont {Kiskis}}, \bibinfo {author} {\bibfnamefont
  {M.}~\bibnamefont {Lin}}, \bibinfo {author} {\bibfnamefont {E.}~\bibnamefont
  {Neil}}, \bibinfo {author} {\bibfnamefont {J.}~\bibnamefont {Osborn}},
  \bibinfo {author} {\bibfnamefont {C.}~\bibnamefont {Rebbi}},  \emph
  {et~al.},\ }\href {\doibase 10.1103/PhysRevLett.112.111601} {\bibfield
  {journal} {\bibinfo  {journal} {Phys. Rev. Lett.}\ }\textbf {\bibinfo
  {volume} {112}},\ \bibinfo {pages} {111601} (\bibinfo {year} {2014})},\
  \Eprint {http://arxiv.org/abs/1311.4889} {arXiv:1311.4889 [hep-ph]}
  \BibitemShut {NoStop}%
%%CITATION = ARXIV:1311.4889;%%
\bibitem [{\citenamefont {Leino}\ \emph
  {et~al.}(2016{\natexlab{a}})\citenamefont {Leino}, \citenamefont
  {Rantalaiho}, \citenamefont {Rummukainen}, \citenamefont {Suorsa},
  \citenamefont {Tuominen},\ and\ \citenamefont {Tähtinen}}]{Leino:2016njf}%
  \BibitemOpen
  \bibfield  {author} {\bibinfo {author} {\bibfnamefont {V.}~\bibnamefont
  {Leino}}, \bibinfo {author} {\bibfnamefont {T.}~\bibnamefont {Rantalaiho}},
  \bibinfo {author} {\bibfnamefont {K.}~\bibnamefont {Rummukainen}}, \bibinfo
  {author} {\bibfnamefont {J.~M.}\ \bibnamefont {Suorsa}}, \bibinfo {author}
  {\bibfnamefont {K.}~\bibnamefont {Tuominen}}, \ and\ \bibinfo {author}
  {\bibfnamefont {S.}~\bibnamefont {Tähtinen}},\ }in\ \href
  {https://inspirehep.net/record/1495256/files/arXiv:1610.09989.pdf} {\emph
  {\bibinfo {booktitle} {{Proceedings, 34th International Symposium on Lattice
  Field Theory (Lattice 2016): Southampton, UK, July 24-30, 2016}}}}\ (\bibinfo
  {year} {2016})\ \Eprint {http://arxiv.org/abs/1610.09989} {arXiv:1610.09989
  [hep-lat]} \BibitemShut {NoStop}%
%%CITATION = ARXIV:1610.09989;%%
\bibitem [{\citenamefont {Fritzsch}\ and\ \citenamefont
  {Ramos}(2013)}]{Fritzsch:2013je}%
  \BibitemOpen
  \bibfield  {author} {\bibinfo {author} {\bibfnamefont {P.}~\bibnamefont
  {Fritzsch}}\ and\ \bibinfo {author} {\bibfnamefont {A.}~\bibnamefont
  {Ramos}},\ }\href {\doibase 10.1007/JHEP10(2013)008} {\bibfield  {journal}
  {\bibinfo  {journal} {JHEP}\ }\textbf {\bibinfo {volume} {10}},\ \bibinfo
  {pages} {008} (\bibinfo {year} {2013})},\ \Eprint
  {http://arxiv.org/abs/1301.4388} {arXiv:1301.4388 [hep-lat]} \BibitemShut
  {NoStop}%
%%CITATION = ARXIV:1301.4388;%%
\bibitem [{\citenamefont {Rantaharju}\ \emph {et~al.}(2014)\citenamefont
  {Rantaharju}, \citenamefont {Karavirta}, \citenamefont {Leino}, \citenamefont
  {Rantalaiho}, \citenamefont {Rummukainen},\ and\ \citenamefont
  {Tuominen}}]{Rantaharju:2014ila}%
  \BibitemOpen
  \bibfield  {author} {\bibinfo {author} {\bibfnamefont {J.}~\bibnamefont
  {Rantaharju}}, \bibinfo {author} {\bibfnamefont {T.}~\bibnamefont
  {Karavirta}}, \bibinfo {author} {\bibfnamefont {V.}~\bibnamefont {Leino}},
  \bibinfo {author} {\bibfnamefont {T.}~\bibnamefont {Rantalaiho}}, \bibinfo
  {author} {\bibfnamefont {K.}~\bibnamefont {Rummukainen}}, \ and\ \bibinfo
  {author} {\bibfnamefont {K.}~\bibnamefont {Tuominen}},\ }\bibfield
  {booktitle} {\emph {\bibinfo {booktitle} {{Proceedings, 32nd International
  Symposium on Lattice Field Theory (Lattice 2014)}}},\ }\href@noop {}
  {\bibfield  {journal} {\bibinfo  {journal} {PoS}\ }\textbf {\bibinfo {volume}
  {LATTICE2014}},\ \bibinfo {pages} {258} (\bibinfo {year} {2014})},\ \Eprint
  {http://arxiv.org/abs/1411.4879} {arXiv:1411.4879 [hep-lat]} \BibitemShut
  {NoStop}%
%%CITATION = ARXIV:1411.4879;%%
\bibitem [{\citenamefont {Suorsa}\ \emph
  {et~al.}(2016{\natexlab{a}})\citenamefont {Suorsa}, \citenamefont {Leino},
  \citenamefont {Rantaharju}, \citenamefont {Rantalaiho}, \citenamefont
  {Rummukainen}, \citenamefont {Tuominen},\ and\ \citenamefont
  {Weir}}]{Suorsa:2015hoh}%
  \BibitemOpen
  \bibfield  {author} {\bibinfo {author} {\bibfnamefont {J.~M.}\ \bibnamefont
  {Suorsa}}, \bibinfo {author} {\bibfnamefont {V.}~\bibnamefont {Leino}},
  \bibinfo {author} {\bibfnamefont {J.}~\bibnamefont {Rantaharju}}, \bibinfo
  {author} {\bibfnamefont {T.}~\bibnamefont {Rantalaiho}}, \bibinfo {author}
  {\bibfnamefont {K.}~\bibnamefont {Rummukainen}}, \bibinfo {author}
  {\bibfnamefont {K.}~\bibnamefont {Tuominen}}, \ and\ \bibinfo {author}
  {\bibfnamefont {D.~J.}\ \bibnamefont {Weir}},\ }\bibfield  {booktitle} {\emph
  {\bibinfo {booktitle} {{Proceedings, 33rd International Symposium on Lattice
  Field Theory (Lattice 2015): Kobe, Japan, July 14-18, 2015}}},\ }\href@noop
  {} {\bibfield  {journal} {\bibinfo  {journal} {PoS}\ }\textbf {\bibinfo
  {volume} {LATTICE2015}},\ \bibinfo {pages} {247} (\bibinfo {year}
  {2016}{\natexlab{a}})},\ \Eprint {http://arxiv.org/abs/1511.03468}
  {arXiv:1511.03468 [hep-lat]} \BibitemShut {NoStop}%
%%CITATION = ARXIV:1511.03468;%%
\bibitem [{\citenamefont {Leino}\ \emph
  {et~al.}(2016{\natexlab{b}})\citenamefont {Leino}, \citenamefont {Karavirta},
  \citenamefont {Rantaharju}, \citenamefont {Rantalaiho}, \citenamefont
  {Rummukainen}, \citenamefont {Suorsa},\ and\ \citenamefont
  {Tuominen}}]{Leino:2015bfg}%
  \BibitemOpen
  \bibfield  {author} {\bibinfo {author} {\bibfnamefont {V.}~\bibnamefont
  {Leino}}, \bibinfo {author} {\bibfnamefont {T.}~\bibnamefont {Karavirta}},
  \bibinfo {author} {\bibfnamefont {J.}~\bibnamefont {Rantaharju}}, \bibinfo
  {author} {\bibfnamefont {T.}~\bibnamefont {Rantalaiho}}, \bibinfo {author}
  {\bibfnamefont {K.}~\bibnamefont {Rummukainen}}, \bibinfo {author}
  {\bibfnamefont {J.~M.}\ \bibnamefont {Suorsa}}, \ and\ \bibinfo {author}
  {\bibfnamefont {K.}~\bibnamefont {Tuominen}},\ }\bibfield  {booktitle} {\emph
  {\bibinfo {booktitle} {{Proceedings, 33rd International Symposium on Lattice
  Field Theory (Lattice 2015): Kobe, Japan, July 14-18, 2015}}},\ }\href@noop
  {} {\bibfield  {journal} {\bibinfo  {journal} {PoS}\ }\textbf {\bibinfo
  {volume} {LATTICE2015}},\ \bibinfo {pages} {226} (\bibinfo {year}
  {2016}{\natexlab{b}})},\ \Eprint {http://arxiv.org/abs/1511.03563}
  {arXiv:1511.03563 [hep-lat]} \BibitemShut {NoStop}%
%%CITATION = ARXIV:1511.03563;%%
\bibitem [{\citenamefont {Suorsa}\ \emph
  {et~al.}(2016{\natexlab{b}})\citenamefont {Suorsa}, \citenamefont {Leino},
  \citenamefont {Rantaharju}, \citenamefont {Rantalaiho}, \citenamefont
  {Rummukainen}, \citenamefont {Tuominen},\ and\ \citenamefont
  {Tähtinen}}]{Suorsa:2016jsf}%
  \BibitemOpen
  \bibfield  {author} {\bibinfo {author} {\bibfnamefont {J.~M.}\ \bibnamefont
  {Suorsa}}, \bibinfo {author} {\bibfnamefont {V.}~\bibnamefont {Leino}},
  \bibinfo {author} {\bibfnamefont {J.}~\bibnamefont {Rantaharju}}, \bibinfo
  {author} {\bibfnamefont {T.}~\bibnamefont {Rantalaiho}}, \bibinfo {author}
  {\bibfnamefont {K.}~\bibnamefont {Rummukainen}}, \bibinfo {author}
  {\bibfnamefont {K.}~\bibnamefont {Tuominen}}, \ and\ \bibinfo {author}
  {\bibfnamefont {S.}~\bibnamefont {Tähtinen}},\ }in\ \href
  {https://inspirehep.net/record/1496007/files/arXiv:1611.02022.pdf} {\emph
  {\bibinfo {booktitle} {{Proceedings, 34th International Symposium on Lattice
  Field Theory (Lattice 2016): Southampton, UK, July 24-30, 2016}}}}\ (\bibinfo
  {year} {2016})\ \Eprint {http://arxiv.org/abs/1611.02022} {arXiv:1611.02022
  [hep-lat]} \BibitemShut {NoStop}%
%%CITATION = ARXIV:1611.02022;%%
\bibitem [{\citenamefont {Capitani}\ \emph {et~al.}(2006)\citenamefont
  {Capitani}, \citenamefont {Durr},\ and\ \citenamefont
  {Hoelbling}}]{Capitani:2006ni}%
  \BibitemOpen
  \bibfield  {author} {\bibinfo {author} {\bibfnamefont {S.}~\bibnamefont
  {Capitani}}, \bibinfo {author} {\bibfnamefont {S.}~\bibnamefont {Durr}}, \
  and\ \bibinfo {author} {\bibfnamefont {C.}~\bibnamefont {Hoelbling}},\ }\href
  {\doibase 10.1088/1126-6708/2006/11/028} {\bibfield  {journal} {\bibinfo
  {journal} {JHEP}\ }\textbf {\bibinfo {volume} {11}},\ \bibinfo {pages} {028}
  (\bibinfo {year} {2006})},\ \Eprint {http://arxiv.org/abs/hep-lat/0607006}
  {arXiv:hep-lat/0607006 [hep-lat]} \BibitemShut {NoStop}%
%%CITATION = HEP-LAT/0607006;%%
\bibitem [{\citenamefont {DeGrand}\ \emph {et~al.}(2011)\citenamefont
  {DeGrand}, \citenamefont {Shamir},\ and\ \citenamefont
  {Svetitsky}}]{DeGrand:2011vp}%
  \BibitemOpen
  \bibfield  {author} {\bibinfo {author} {\bibfnamefont {T.}~\bibnamefont
  {DeGrand}}, \bibinfo {author} {\bibfnamefont {Y.}~\bibnamefont {Shamir}}, \
  and\ \bibinfo {author} {\bibfnamefont {B.}~\bibnamefont {Svetitsky}},\
  }\bibfield  {booktitle} {\emph {\bibinfo {booktitle} {{Proceedings, 29th
  International Symposium on Lattice field theory (Lattice 2011): Squaw Valley,
  Lake Tahoe, USA, July 10-16, 2011}}},\ }\href@noop {} {\bibfield  {journal}
  {\bibinfo  {journal} {PoS}\ }\textbf {\bibinfo {volume} {LATTICE2011}},\
  \bibinfo {pages} {060} (\bibinfo {year} {2011})},\ \Eprint
  {http://arxiv.org/abs/1110.6845} {arXiv:1110.6845 [hep-lat]} \BibitemShut
  {NoStop}%
%%CITATION = ARXIV:1110.6845;%%
\bibitem [{\citenamefont {Rantaharju}\ \emph {et~al.}(2016)\citenamefont
  {Rantaharju}, \citenamefont {Rantalaiho}, \citenamefont {Rummukainen},\ and\
  \citenamefont {Tuominen}}]{Rantaharju:2015yva}%
  \BibitemOpen
  \bibfield  {author} {\bibinfo {author} {\bibfnamefont {J.}~\bibnamefont
  {Rantaharju}}, \bibinfo {author} {\bibfnamefont {T.}~\bibnamefont
  {Rantalaiho}}, \bibinfo {author} {\bibfnamefont {K.}~\bibnamefont
  {Rummukainen}}, \ and\ \bibinfo {author} {\bibfnamefont {K.}~\bibnamefont
  {Tuominen}},\ }\href {\doibase 10.1103/PhysRevD.93.094509} {\bibfield
  {journal} {\bibinfo  {journal} {Phys. Rev.}\ }\textbf {\bibinfo {volume}
  {D93}},\ \bibinfo {pages} {094509} (\bibinfo {year} {2016})},\ \Eprint
  {http://arxiv.org/abs/1510.03335} {arXiv:1510.03335 [hep-lat]} \BibitemShut
  {NoStop}%
%%CITATION = ARXIV:1510.03335;%%
\bibitem [{\citenamefont {Luscher}\ \emph {et~al.}(1992)\citenamefont
  {Luscher}, \citenamefont {Narayanan}, \citenamefont {Weisz},\ and\
  \citenamefont {Wolff}}]{Luscher:1992an}%
  \BibitemOpen
  \bibfield  {author} {\bibinfo {author} {\bibfnamefont {M.}~\bibnamefont
  {Luscher}}, \bibinfo {author} {\bibfnamefont {R.}~\bibnamefont {Narayanan}},
  \bibinfo {author} {\bibfnamefont {P.}~\bibnamefont {Weisz}}, \ and\ \bibinfo
  {author} {\bibfnamefont {U.}~\bibnamefont {Wolff}},\ }\href {\doibase
  10.1016/0550-3213(92)90466-O} {\bibfield  {journal} {\bibinfo  {journal}
  {Nucl. Phys.}\ }\textbf {\bibinfo {volume} {B384}},\ \bibinfo {pages} {168}
  (\bibinfo {year} {1992})},\ \Eprint {http://arxiv.org/abs/hep-lat/9207009}
  {arXiv:hep-lat/9207009 [hep-lat]} \BibitemShut {NoStop}%
%%CITATION = HEP-LAT/9207009;%%
\bibitem [{\citenamefont {Luscher}\ \emph {et~al.}(1993)\citenamefont
  {Luscher}, \citenamefont {Narayanan}, \citenamefont {Sommer}, \citenamefont
  {Wolff},\ and\ \citenamefont {Weisz}}]{Luscher:1992ny}%
  \BibitemOpen
  \bibfield  {author} {\bibinfo {author} {\bibfnamefont {M.}~\bibnamefont
  {Luscher}}, \bibinfo {author} {\bibfnamefont {R.}~\bibnamefont {Narayanan}},
  \bibinfo {author} {\bibfnamefont {R.}~\bibnamefont {Sommer}}, \bibinfo
  {author} {\bibfnamefont {U.}~\bibnamefont {Wolff}}, \ and\ \bibinfo {author}
  {\bibfnamefont {P.}~\bibnamefont {Weisz}},\ }\bibfield  {booktitle} {\emph
  {\bibinfo {booktitle} {{The International Symposium on Lattice Field Theory:
  Lattice 92 Amsterdam, Netherlands, September 15-19, 1992}}},\ }\href
  {\doibase 10.1016/0920-5632(93)90183-7} {\bibfield  {journal} {\bibinfo
  {journal} {Nucl. Phys. Proc. Suppl.}\ }\textbf {\bibinfo {volume} {30}},\
  \bibinfo {pages} {139} (\bibinfo {year} {1993})}\BibitemShut {NoStop}%
%%CITATION = NUPHZ,30,139;%%
\bibitem [{\citenamefont {Luscher}\ \emph {et~al.}(1994)\citenamefont
  {Luscher}, \citenamefont {Sommer}, \citenamefont {Weisz},\ and\ \citenamefont
  {Wolff}}]{Luscher:1993gh}%
  \BibitemOpen
  \bibfield  {author} {\bibinfo {author} {\bibfnamefont {M.}~\bibnamefont
  {Luscher}}, \bibinfo {author} {\bibfnamefont {R.}~\bibnamefont {Sommer}},
  \bibinfo {author} {\bibfnamefont {P.}~\bibnamefont {Weisz}}, \ and\ \bibinfo
  {author} {\bibfnamefont {U.}~\bibnamefont {Wolff}},\ }\href {\doibase
  10.1016/0550-3213(94)90629-7} {\bibfield  {journal} {\bibinfo  {journal}
  {Nucl. Phys.}\ }\textbf {\bibinfo {volume} {B413}},\ \bibinfo {pages} {481}
  (\bibinfo {year} {1994})},\ \Eprint {http://arxiv.org/abs/hep-lat/9309005}
  {arXiv:hep-lat/9309005 [hep-lat]} \BibitemShut {NoStop}%
%%CITATION = HEP-LAT/9309005;%%
\bibitem [{\citenamefont {Della~Morte}\ \emph
  {et~al.}(2005{\natexlab{a}})\citenamefont {Della~Morte}, \citenamefont
  {Frezzotti}, \citenamefont {Heitger}, \citenamefont {Rolf}, \citenamefont
  {Sommer},\ and\ \citenamefont {Wolff}}]{DellaMorte:2004bc}%
  \BibitemOpen
  \bibfield  {author} {\bibinfo {author} {\bibfnamefont {M.}~\bibnamefont
  {Della~Morte}}, \bibinfo {author} {\bibfnamefont {R.}~\bibnamefont
  {Frezzotti}}, \bibinfo {author} {\bibfnamefont {J.}~\bibnamefont {Heitger}},
  \bibinfo {author} {\bibfnamefont {J.}~\bibnamefont {Rolf}}, \bibinfo {author}
  {\bibfnamefont {R.}~\bibnamefont {Sommer}}, \ and\ \bibinfo {author}
  {\bibfnamefont {U.}~\bibnamefont {Wolff}} (\bibinfo {collaboration}
  {ALPHA}),\ }\href {\doibase 10.1016/j.nuclphysb.2005.02.013} {\bibfield
  {journal} {\bibinfo  {journal} {Nucl. Phys.}\ }\textbf {\bibinfo {volume}
  {B713}},\ \bibinfo {pages} {378} (\bibinfo {year} {2005}{\natexlab{a}})},\
  \Eprint {http://arxiv.org/abs/hep-lat/0411025} {arXiv:hep-lat/0411025
  [hep-lat]} \BibitemShut {NoStop}%
%%CITATION = HEP-LAT/0411025;%%
\bibitem [{\citenamefont {Luscher}\ and\ \citenamefont
  {Weisz}(1996)}]{Luscher:1996vw}%
  \BibitemOpen
  \bibfield  {author} {\bibinfo {author} {\bibfnamefont {M.}~\bibnamefont
  {Luscher}}\ and\ \bibinfo {author} {\bibfnamefont {P.}~\bibnamefont
  {Weisz}},\ }\href {\doibase 10.1016/0550-3213(96)00448-8} {\bibfield
  {journal} {\bibinfo  {journal} {Nucl. Phys.}\ }\textbf {\bibinfo {volume}
  {B479}},\ \bibinfo {pages} {429} (\bibinfo {year} {1996})},\ \Eprint
  {http://arxiv.org/abs/hep-lat/9606016} {arXiv:hep-lat/9606016 [hep-lat]}
  \BibitemShut {NoStop}%
%%CITATION = HEP-LAT/9606016;%%
\bibitem [{\citenamefont {Narayanan}\ and\ \citenamefont
  {Neuberger}(2006)}]{Narayanan:2006rf}%
  \BibitemOpen
  \bibfield  {author} {\bibinfo {author} {\bibfnamefont {R.}~\bibnamefont
  {Narayanan}}\ and\ \bibinfo {author} {\bibfnamefont {H.}~\bibnamefont
  {Neuberger}},\ }\href {\doibase 10.1088/1126-6708/2006/03/064} {\bibfield
  {journal} {\bibinfo  {journal} {JHEP}\ }\textbf {\bibinfo {volume} {03}},\
  \bibinfo {pages} {064} (\bibinfo {year} {2006})},\ \Eprint
  {http://arxiv.org/abs/hep-th/0601210} {arXiv:hep-th/0601210 [hep-th]}
  \BibitemShut {NoStop}%
%%CITATION = HEP-TH/0601210;%%
\bibitem [{\citenamefont {Luscher}(2010)}]{Luscher:2009eq}%
  \BibitemOpen
  \bibfield  {author} {\bibinfo {author} {\bibfnamefont {M.}~\bibnamefont
  {Luscher}},\ }\href {\doibase 10.1007/s00220-009-0953-7} {\bibfield
  {journal} {\bibinfo  {journal} {Commun. Math. Phys.}\ }\textbf {\bibinfo
  {volume} {293}},\ \bibinfo {pages} {899} (\bibinfo {year} {2010})},\ \Eprint
  {http://arxiv.org/abs/0907.5491} {arXiv:0907.5491 [hep-lat]} \BibitemShut
  {NoStop}%
%%CITATION = ARXIV:0907.5491;%%
\bibitem [{\citenamefont {Lüscher}(2010)}]{Luscher:2010iy}%
  \BibitemOpen
  \bibfield  {author} {\bibinfo {author} {\bibfnamefont {M.}~\bibnamefont
  {Lüscher}},\ }\href {\doibase 10.1007/JHEP08(2010)071,
  10.1007/JHEP03(2014)092} {\bibfield  {journal} {\bibinfo  {journal} {JHEP}\
  }\textbf {\bibinfo {volume} {08}},\ \bibinfo {pages} {071} (\bibinfo {year}
  {2010})},\ \bibinfo {note} {[Erratum: JHEP03,092(2014)]},\ \Eprint
  {http://arxiv.org/abs/1006.4518} {arXiv:1006.4518 [hep-lat]} \BibitemShut
  {NoStop}%
%%CITATION = ARXIV:1006.4518;%%
\bibitem [{\citenamefont {Luscher}\ and\ \citenamefont
  {Weisz}(2011)}]{Luscher:2011bx}%
  \BibitemOpen
  \bibfield  {author} {\bibinfo {author} {\bibfnamefont {M.}~\bibnamefont
  {Luscher}}\ and\ \bibinfo {author} {\bibfnamefont {P.}~\bibnamefont
  {Weisz}},\ }\href {\doibase 10.1007/JHEP02(2011)051} {\bibfield  {journal}
  {\bibinfo  {journal} {JHEP}\ }\textbf {\bibinfo {volume} {02}},\ \bibinfo
  {pages} {051} (\bibinfo {year} {2011})},\ \Eprint
  {http://arxiv.org/abs/1101.0963} {arXiv:1101.0963 [hep-th]} \BibitemShut
  {NoStop}%
%%CITATION = ARXIV:1101.0963;%%
\bibitem [{\citenamefont {Luscher}\ and\ \citenamefont
  {Weisz}(1985)}]{Luscher:1984xn}%
  \BibitemOpen
  \bibfield  {author} {\bibinfo {author} {\bibfnamefont {M.}~\bibnamefont
  {Luscher}}\ and\ \bibinfo {author} {\bibfnamefont {P.}~\bibnamefont
  {Weisz}},\ }\href {\doibase 10.1007/BF01206178} {\bibfield  {journal}
  {\bibinfo  {journal} {Commun. Math. Phys.}\ }\textbf {\bibinfo {volume}
  {97}},\ \bibinfo {pages} {59} (\bibinfo {year} {1985})},\ \bibinfo {note}
  {[Erratum: Commun. Math. Phys.98,433(1985)]}\BibitemShut {NoStop}%
%%CITATION = CMPHA,97,59;%%
\bibitem [{\citenamefont {Fodor}\ \emph
  {et~al.}(2012{\natexlab{a}})\citenamefont {Fodor}, \citenamefont {Holland},
  \citenamefont {Kuti}, \citenamefont {Nogradi},\ and\ \citenamefont
  {Wong}}]{Fodor:2012td}%
  \BibitemOpen
  \bibfield  {author} {\bibinfo {author} {\bibfnamefont {Z.}~\bibnamefont
  {Fodor}}, \bibinfo {author} {\bibfnamefont {K.}~\bibnamefont {Holland}},
  \bibinfo {author} {\bibfnamefont {J.}~\bibnamefont {Kuti}}, \bibinfo {author}
  {\bibfnamefont {D.}~\bibnamefont {Nogradi}}, \ and\ \bibinfo {author}
  {\bibfnamefont {C.~H.}\ \bibnamefont {Wong}},\ }\href {\doibase
  10.1007/JHEP11(2012)007} {\bibfield  {journal} {\bibinfo  {journal} {JHEP}\
  }\textbf {\bibinfo {volume} {11}},\ \bibinfo {pages} {007} (\bibinfo {year}
  {2012}{\natexlab{a}})},\ \Eprint {http://arxiv.org/abs/1208.1051}
  {arXiv:1208.1051 [hep-lat]} \BibitemShut {NoStop}%
%%CITATION = ARXIV:1208.1051;%%
\bibitem [{\citenamefont {Fodor}\ \emph
  {et~al.}(2012{\natexlab{b}})\citenamefont {Fodor}, \citenamefont {Holland},
  \citenamefont {Kuti}, \citenamefont {Nogradi},\ and\ \citenamefont
  {Wong}}]{Fodor:2012qh}%
  \BibitemOpen
  \bibfield  {author} {\bibinfo {author} {\bibfnamefont {Z.}~\bibnamefont
  {Fodor}}, \bibinfo {author} {\bibfnamefont {K.}~\bibnamefont {Holland}},
  \bibinfo {author} {\bibfnamefont {J.}~\bibnamefont {Kuti}}, \bibinfo {author}
  {\bibfnamefont {D.}~\bibnamefont {Nogradi}}, \ and\ \bibinfo {author}
  {\bibfnamefont {C.~H.}\ \bibnamefont {Wong}},\ }\bibfield  {booktitle} {\emph
  {\bibinfo {booktitle} {{Proceedings, 30th International Symposium on Lattice
  Field Theory (Lattice 2012)}}},\ }\href@noop {} {\bibfield  {journal}
  {\bibinfo  {journal} {PoS}\ }\textbf {\bibinfo {volume} {LATTICE2012}},\
  \bibinfo {pages} {050} (\bibinfo {year} {2012}{\natexlab{b}})},\ \Eprint
  {http://arxiv.org/abs/1211.3247} {arXiv:1211.3247 [hep-lat]} \BibitemShut
  {NoStop}%
%%CITATION = ARXIV:1211.3247;%%
\bibitem [{\citenamefont {Lüscher}(2014)}]{Luscher:2014kea}%
  \BibitemOpen
  \bibfield  {author} {\bibinfo {author} {\bibfnamefont {M.}~\bibnamefont
  {Lüscher}},\ }\href {\doibase 10.1007/JHEP06(2014)105} {\bibfield  {journal}
  {\bibinfo  {journal} {JHEP}\ }\textbf {\bibinfo {volume} {06}},\ \bibinfo
  {pages} {105} (\bibinfo {year} {2014})},\ \Eprint
  {http://arxiv.org/abs/1404.5930} {arXiv:1404.5930 [hep-lat]} \BibitemShut
  {NoStop}%
%%CITATION = ARXIV:1404.5930;%%
\bibitem [{\citenamefont {Ramos}\ and\ \citenamefont
  {Sint}(2016)}]{Ramos:2015baa}%
  \BibitemOpen
  \bibfield  {author} {\bibinfo {author} {\bibfnamefont {A.}~\bibnamefont
  {Ramos}}\ and\ \bibinfo {author} {\bibfnamefont {S.}~\bibnamefont {Sint}},\
  }\href {\doibase 10.1140/epjc/s10052-015-3831-9} {\bibfield  {journal}
  {\bibinfo  {journal} {Eur. Phys. J.}\ }\textbf {\bibinfo {volume} {C76}},\
  \bibinfo {pages} {15} (\bibinfo {year} {2016})},\ \Eprint
  {http://arxiv.org/abs/1508.05552} {arXiv:1508.05552 [hep-lat]} \BibitemShut
  {NoStop}%
%%CITATION = ARXIV:1508.05552;%%
\bibitem [{\citenamefont {Cheng}\ \emph {et~al.}(2014)\citenamefont {Cheng},
  \citenamefont {Hasenfratz}, \citenamefont {Liu}, \citenamefont
  {Petropoulos},\ and\ \citenamefont {Schaich}}]{Cheng:2014jba}%
  \BibitemOpen
  \bibfield  {author} {\bibinfo {author} {\bibfnamefont {A.}~\bibnamefont
  {Cheng}}, \bibinfo {author} {\bibfnamefont {A.}~\bibnamefont {Hasenfratz}},
  \bibinfo {author} {\bibfnamefont {Y.}~\bibnamefont {Liu}}, \bibinfo {author}
  {\bibfnamefont {G.}~\bibnamefont {Petropoulos}}, \ and\ \bibinfo {author}
  {\bibfnamefont {D.}~\bibnamefont {Schaich}},\ }\href {\doibase
  10.1007/JHEP05(2014)137} {\bibfield  {journal} {\bibinfo  {journal} {JHEP}\
  }\textbf {\bibinfo {volume} {05}},\ \bibinfo {pages} {137} (\bibinfo {year}
  {2014})},\ \Eprint {http://arxiv.org/abs/1404.0984} {arXiv:1404.0984
  [hep-lat]} \BibitemShut {NoStop}%
%%CITATION = ARXIV:1404.0984;%%
\bibitem [{\citenamefont {Sint}\ and\ \citenamefont
  {Weisz}(1999)}]{Sint:1998iq}%
  \BibitemOpen
  \bibfield  {author} {\bibinfo {author} {\bibfnamefont {S.}~\bibnamefont
  {Sint}}\ and\ \bibinfo {author} {\bibfnamefont {P.}~\bibnamefont {Weisz}}
  (\bibinfo {collaboration} {ALPHA}),\ }\href {\doibase
  10.1016/S0550-3213(98)00874-8} {\bibfield  {journal} {\bibinfo  {journal}
  {Nucl. Phys.}\ }\textbf {\bibinfo {volume} {B545}},\ \bibinfo {pages} {529}
  (\bibinfo {year} {1999})},\ \Eprint {http://arxiv.org/abs/hep-lat/9808013}
  {arXiv:hep-lat/9808013 [hep-lat]} \BibitemShut {NoStop}%
%%CITATION = HEP-LAT/9808013;%%
\bibitem [{\citenamefont {Capitani}\ \emph {et~al.}(1999)\citenamefont
  {Capitani}, \citenamefont {Luscher}, \citenamefont {Sommer},\ and\
  \citenamefont {Wittig}}]{Capitani:1998mq}%
  \BibitemOpen
  \bibfield  {author} {\bibinfo {author} {\bibfnamefont {S.}~\bibnamefont
  {Capitani}}, \bibinfo {author} {\bibfnamefont {M.}~\bibnamefont {Luscher}},
  \bibinfo {author} {\bibfnamefont {R.}~\bibnamefont {Sommer}}, \ and\ \bibinfo
  {author} {\bibfnamefont {H.}~\bibnamefont {Wittig}} (\bibinfo {collaboration}
  {ALPHA}),\ }\href {\doibase 10.1016/S0550-3213(98)00857-8} {\bibfield
  {journal} {\bibinfo  {journal} {Nucl. Phys.}\ }\textbf {\bibinfo {volume}
  {B544}},\ \bibinfo {pages} {669} (\bibinfo {year} {1999})},\ \Eprint
  {http://arxiv.org/abs/hep-lat/9810063} {arXiv:hep-lat/9810063 [hep-lat]}
  \BibitemShut {NoStop}%
%%CITATION = HEP-LAT/9810063;%%
\bibitem [{\citenamefont {Della~Morte}\ \emph
  {et~al.}(2005{\natexlab{b}})\citenamefont {Della~Morte}, \citenamefont
  {Hoffmann}, \citenamefont {Knechtli}, \citenamefont {Rolf}, \citenamefont
  {Sommer}, \citenamefont {Wetzorke},\ and\ \citenamefont
  {Wolff}}]{DellaMorte:2005kg}%
  \BibitemOpen
  \bibfield  {author} {\bibinfo {author} {\bibfnamefont {M.}~\bibnamefont
  {Della~Morte}}, \bibinfo {author} {\bibfnamefont {R.}~\bibnamefont
  {Hoffmann}}, \bibinfo {author} {\bibfnamefont {F.}~\bibnamefont {Knechtli}},
  \bibinfo {author} {\bibfnamefont {J.}~\bibnamefont {Rolf}}, \bibinfo {author}
  {\bibfnamefont {R.}~\bibnamefont {Sommer}}, \bibinfo {author} {\bibfnamefont
  {I.}~\bibnamefont {Wetzorke}}, \ and\ \bibinfo {author} {\bibfnamefont
  {U.}~\bibnamefont {Wolff}} (\bibinfo {collaboration} {ALPHA}),\ }\href
  {\doibase 10.1016/j.nuclphysb.2005.09.028} {\bibfield  {journal} {\bibinfo
  {journal} {Nucl. Phys.}\ }\textbf {\bibinfo {volume} {B729}},\ \bibinfo
  {pages} {117} (\bibinfo {year} {2005}{\natexlab{b}})},\ \Eprint
  {http://arxiv.org/abs/hep-lat/0507035} {arXiv:hep-lat/0507035 [hep-lat]}
  \BibitemShut {NoStop}%
%%CITATION = HEP-LAT/0507035;%%
\bibitem [{\citenamefont {Del~Debbio}\ and\ \citenamefont
  {Zwicky}(2010)}]{DelDebbio:2010ze}%
  \BibitemOpen
  \bibfield  {author} {\bibinfo {author} {\bibfnamefont {L.}~\bibnamefont
  {Del~Debbio}}\ and\ \bibinfo {author} {\bibfnamefont {R.}~\bibnamefont
  {Zwicky}},\ }\href {\doibase 10.1103/PhysRevD.82.014502} {\bibfield
  {journal} {\bibinfo  {journal} {Phys. Rev.}\ }\textbf {\bibinfo {volume}
  {D82}},\ \bibinfo {pages} {014502} (\bibinfo {year} {2010})},\ \Eprint
  {http://arxiv.org/abs/1005.2371} {arXiv:1005.2371 [hep-ph]} \BibitemShut
  {NoStop}%
%%CITATION = ARXIV:1005.2371;%%
\bibitem [{\citenamefont {Giusti}\ and\ \citenamefont
  {Luscher}(2009)}]{Giusti:2008vb}%
  \BibitemOpen
  \bibfield  {author} {\bibinfo {author} {\bibfnamefont {L.}~\bibnamefont
  {Giusti}}\ and\ \bibinfo {author} {\bibfnamefont {M.}~\bibnamefont
  {Luscher}},\ }\href {\doibase 10.1088/1126-6708/2009/03/013} {\bibfield
  {journal} {\bibinfo  {journal} {JHEP}\ }\textbf {\bibinfo {volume} {03}},\
  \bibinfo {pages} {013} (\bibinfo {year} {2009})},\ \Eprint
  {http://arxiv.org/abs/0812.3638} {arXiv:0812.3638 [hep-lat]} \BibitemShut
  {NoStop}%
%%CITATION = ARXIV:0812.3638;%%
\bibitem [{\citenamefont {Patella}(2011)}]{Patella:2011jr}%
  \BibitemOpen
  \bibfield  {author} {\bibinfo {author} {\bibfnamefont {A.}~\bibnamefont
  {Patella}},\ }\href {\doibase 10.1103/PhysRevD.84.125033} {\bibfield
  {journal} {\bibinfo  {journal} {Phys. Rev.}\ }\textbf {\bibinfo {volume}
  {D84}},\ \bibinfo {pages} {125033} (\bibinfo {year} {2011})},\ \Eprint
  {http://arxiv.org/abs/1106.3494} {arXiv:1106.3494 [hep-th]} \BibitemShut
  {NoStop}%
%%CITATION = ARXIV:1106.3494;%%
\bibitem [{\citenamefont {Omelyan}\ \emph {et~al.}(2003)\citenamefont
  {Omelyan}, \citenamefont {Mryglod},\ and\ \citenamefont
  {Folk}}]{Omelyan:2003:SAI}%
  \BibitemOpen
  \bibfield  {author} {\bibinfo {author} {\bibfnamefont {I.~P.}\ \bibnamefont
  {Omelyan}}, \bibinfo {author} {\bibfnamefont {I.~M.}\ \bibnamefont
  {Mryglod}}, \ and\ \bibinfo {author} {\bibfnamefont {R.}~\bibnamefont
  {Folk}},\ }\href {\doibase http://dx.doi.org/10.1016/S0010-4655(02)00754-3}
  {\bibfield  {journal} {\bibinfo  {journal} {{Computer Physics
  Communications}}\ }\textbf {\bibinfo {volume} {151}},\ \bibinfo {pages} {272}
  (\bibinfo {year} {2003})}\BibitemShut {NoStop}%
\bibitem [{\citenamefont {Takaishi}\ and\ \citenamefont
  {de~Forcrand}(2006)}]{Takaishi:2005tz}%
  \BibitemOpen
  \bibfield  {author} {\bibinfo {author} {\bibfnamefont {T.}~\bibnamefont
  {Takaishi}}\ and\ \bibinfo {author} {\bibfnamefont {P.}~\bibnamefont
  {de~Forcrand}},\ }\href {\doibase 10.1103/PhysRevE.73.036706} {\bibfield
  {journal} {\bibinfo  {journal} {Phys. Rev.}\ }\textbf {\bibinfo {volume}
  {E73}},\ \bibinfo {pages} {036706} (\bibinfo {year} {2006})},\ \Eprint
  {http://arxiv.org/abs/hep-lat/0505020} {arXiv:hep-lat/0505020 [hep-lat]}
  \BibitemShut {NoStop}%
%%CITATION = HEP-LAT/0505020;%%
\bibitem [{\citenamefont {Brower}\ \emph {et~al.}(1997)\citenamefont {Brower},
  \citenamefont {Ivanenko}, \citenamefont {Levi},\ and\ \citenamefont
  {Orginos}}]{Brower:1995vx}%
  \BibitemOpen
  \bibfield  {author} {\bibinfo {author} {\bibfnamefont {R.~C.}\ \bibnamefont
  {Brower}}, \bibinfo {author} {\bibfnamefont {T.}~\bibnamefont {Ivanenko}},
  \bibinfo {author} {\bibfnamefont {A.~R.}\ \bibnamefont {Levi}}, \ and\
  \bibinfo {author} {\bibfnamefont {K.~N.}\ \bibnamefont {Orginos}},\ }\href
  {\doibase 10.1016/S0550-3213(96)00579-2} {\bibfield  {journal} {\bibinfo
  {journal} {Nucl. Phys.}\ }\textbf {\bibinfo {volume} {B484}},\ \bibinfo
  {pages} {353} (\bibinfo {year} {1997})},\ \Eprint
  {http://arxiv.org/abs/hep-lat/9509012} {arXiv:hep-lat/9509012 [hep-lat]}
  \BibitemShut {NoStop}%
%%CITATION = HEP-LAT/9509012;%%
\bibitem [{\citenamefont {Hasenfratz}\ \emph
  {et~al.}(2015{\natexlab{a}})\citenamefont {Hasenfratz}, \citenamefont
  {Schaich},\ and\ \citenamefont {Veernala}}]{Hasenfratz:2014rna}%
  \BibitemOpen
  \bibfield  {author} {\bibinfo {author} {\bibfnamefont {A.}~\bibnamefont
  {Hasenfratz}}, \bibinfo {author} {\bibfnamefont {D.}~\bibnamefont {Schaich}},
  \ and\ \bibinfo {author} {\bibfnamefont {A.}~\bibnamefont {Veernala}},\
  }\href {\doibase 10.1007/JHEP06(2015)143} {\bibfield  {journal} {\bibinfo
  {journal} {JHEP}\ }\textbf {\bibinfo {volume} {06}},\ \bibinfo {pages} {143}
  (\bibinfo {year} {2015}{\natexlab{a}})},\ \Eprint
  {http://arxiv.org/abs/1410.5886} {arXiv:1410.5886 [hep-lat]} \BibitemShut
  {NoStop}%
%%CITATION = ARXIV:1410.5886;%%
\bibitem [{\citenamefont {Ramos}(2015)}]{Ramos:2015dla}%
  \BibitemOpen
  \bibfield  {author} {\bibinfo {author} {\bibfnamefont {A.}~\bibnamefont
  {Ramos}},\ }\bibfield  {booktitle} {\emph {\bibinfo {booktitle}
  {{Proceedings, 32nd International Symposium on Lattice Field Theory (Lattice
  2014): Brookhaven, NY, USA, June 23-28, 2014}}},\ }\href@noop {} {\bibfield
  {journal} {\bibinfo  {journal} {PoS}\ }\textbf {\bibinfo {volume}
  {LATTICE2014}},\ \bibinfo {pages} {017} (\bibinfo {year} {2015})},\ \Eprint
  {http://arxiv.org/abs/1506.00118} {arXiv:1506.00118 [hep-lat]} \BibitemShut
  {NoStop}%
%%CITATION = ARXIV:1506.00118;%%
\bibitem [{\citenamefont {Lin}\ \emph {et~al.}(2015)\citenamefont {Lin},
  \citenamefont {Ogawa},\ and\ \citenamefont {Ramos}}]{Lin:2015zpa}%
  \BibitemOpen
  \bibfield  {author} {\bibinfo {author} {\bibfnamefont {C.~J.~D.}\
  \bibnamefont {Lin}}, \bibinfo {author} {\bibfnamefont {K.}~\bibnamefont
  {Ogawa}}, \ and\ \bibinfo {author} {\bibfnamefont {A.}~\bibnamefont
  {Ramos}},\ }\href {\doibase 10.1007/JHEP12(2015)103} {\bibfield  {journal}
  {\bibinfo  {journal} {JHEP}\ }\textbf {\bibinfo {volume} {12}},\ \bibinfo
  {pages} {103} (\bibinfo {year} {2015})},\ \Eprint
  {http://arxiv.org/abs/1510.05755} {arXiv:1510.05755 [hep-lat]} \BibitemShut
  {NoStop}%
%%CITATION = ARXIV:1510.05755;%%
\bibitem [{\citenamefont {Hasenfratz}\ \emph
  {et~al.}(2015{\natexlab{b}})\citenamefont {Hasenfratz}, \citenamefont {Liu},\
  and\ \citenamefont {Huang}}]{Hasenfratz:2015ssa}%
  \BibitemOpen
  \bibfield  {author} {\bibinfo {author} {\bibfnamefont {A.}~\bibnamefont
  {Hasenfratz}}, \bibinfo {author} {\bibfnamefont {Y.}~\bibnamefont {Liu}}, \
  and\ \bibinfo {author} {\bibfnamefont {C.~Y.-H.}\ \bibnamefont {Huang}},\
  }\href@noop {} {\  (\bibinfo {year} {2015}{\natexlab{b}})},\ \Eprint
  {http://arxiv.org/abs/1507.08260} {arXiv:1507.08260 [hep-lat]} \BibitemShut
  {NoStop}%
%%CITATION = ARXIV:1507.08260;%%
\bibitem [{MIL()}]{MILC}%
  \BibitemOpen
  \href@noop {} {\enquote {\bibinfo {title}
  {http://physics.utah.edu/$\sim$detar/milc.html},}\ }\BibitemShut {NoStop}%
\end{thebibliography}%
\bibliographystyle{apsrev4-1.bst}

\end{document}